\documentclass[11pt]{article}
\usepackage{graphicx}
\usepackage{amssymb,amsmath}
\usepackage{xcolor}
\usepackage[sort&compress,numbers]{natbib}

\begin{document}

\title{
\textbf{Corrections to finite--size scaling in \\ the $\varphi^4$ model on square lattices}
}

\author{J. Kaupu\v{z}s$^{1,2}$ 
\thanks{E--mail: \texttt{kaupuzs@latnet.lv}} \hspace{1ex}, 
R. V. N. Melnik$^3$, 
J. Rim\v{s}\=ans$^{1,2,3}$ \\
$^1$Institute of Mathematics and Computer Science, University of Latvia\\
29 Rai\c{n}a Boulevard, LV--1459 Riga, Latvia \\
$^2$ Institute of Mathematical Sciences and Information Technologies, \\
University of Liepaja, 14 Liela Street, Liepaja LV--3401, Latvia \\
$^3$ The MS2 Discovery Interdisciplinary Research Institute, \\
Wilfrid Laurier University, Waterloo, Ontario, Canada, N2L 3C5}

\maketitle

\begin{abstract}
Corrections to scaling in the two--dimensional scalar $\varphi^4$ model are studied
based on non--perturbative analytical arguments and Monte Carlo (MC) simulation data for
different lattice sizes $L$  ($4 \le L \le 1536$) and different values of the $\varphi^4$ coupling constant
$\lambda$, i.~e., $\lambda = 0.1, 1, 10$. 
According to our analysis, amplitudes of the nontrivial correction 
terms with the correction--to--scaling exponents $\omega_{\ell} < 1$ become small when approaching the Ising limit ($\lambda \to \infty$),
but such corrections generally exist in the 2D $\varphi^4$ model. Analytical arguments show the existence of corrections
with the exponent $3/4$. The numerical analysis suggests that there exist also
corrections with the exponent $1/2$ and, very likely, also corrections with the exponent about $1/4$, which
are detectable at $\lambda = 0.1$.
The numerical tests clearly show that the structure of corrections to scaling
in the 2D $\varphi^4$ model 
differs from the usually expected one in the 2D Ising model.
\end{abstract}


\section{Introduction}
\label{intro}

The $\varphi^4$ model is one of the most extensively used tools in analytical studies
of critical phenomena -- see, e.~g.,\cite{Amit,Ma,Justin,Kleinert,PV,K_Ann01,K2012x}. These studies have risen also
a significant interest in numerical testing of the theoretical results for this model. 
Recently, some challenging non--perturbative analytical results for the corrections to scaling in the
$\varphi^4$ model have been obtained~\cite{K2012}, which could be relatively easily verified numerically in the
two--dimensional case. Therefore, we will further focus just on this case.
Although the analytical studies are based on the continuous $\varphi^4$ model, its lattice version
is more convenient for Monte Carlo (MC) simulations. Earlier MC studies of the 2D lattice model go back to the work by
Milchev, Heermann and Binder~\cite{MHB86}. The continuous version has been simulated, e.~g., in~\cite{TC90}.
In~\cite{MHB86}, effective critical exponents $\nu \approx 0.8$ for correlation length
and $\gamma \approx 1.25$ for susceptibility have
been obtained, based on the simulation data for lattices sizes up to $L=20$. The considered
there a scalar 2D $\varphi^4$ model should belong to the 2D Ising universality class
with the exponents $\nu = 1$ and $\gamma = 7/4$, so that these effective exponents point
to the presence of remarkable corrections to scaling. A later MC study~\cite{MF92} of larger
lattices, up to $L=128$, has supported the idea that this model belongs to the 2D Ising universality
class, stating that the asymptotic scaling is achieved for $L \gtrsim 32$. Apparently, numerical studies cause no
doubts that the leading scaling exponents for the two--dimensional scalar $\varphi^4$ model and the 2D Ising model
are the same. However, it is still important to refine further corrections to scaling. 
Indeed, the 2D $\varphi^4$ model can contain nontrivial correction terms, which do not show up or cancel in the 2D Ising model.
We will focus on this issue in the following sections.

\section{Analytical arguments}
\label{sec:analytical}

In~\cite{K2012}, a theorem has been proven concerning corrections to scaling in the continuous $\varphi^4$ model,
based on a set of assumptions, i.~e., certain conditions stated in the theorem. Based on this theorem,
it has been argued in~\cite{K2012} that the two--point correlation function contains a correction term
with the correction--to--scaling exponent $\theta_{\ell} =\gamma-1$ if $\gamma >1$ holds for the susceptibility 
exponent $\gamma$. Here we reconsider these 
non--perturbative analytical arguments by proving a new theorem, leading to the same conclusions
at even better (softer) natural assumptions, which have been verified numerically.

We consider the continuous $\varphi^4$ model in the 
thermodynamic limit of diverging volume $V \to \infty$ with the Hamiltonian $\cal{H}$ given by
\begin{equation} \label{eq:H}
\frac{\cal{H}}{k_B T}= \int \left( r_0 \varphi^2({\bf x}) + c (\nabla \varphi({\bf x}))^2 
+ u \varphi^4({\bf x}) \right) d {\bf x} \;,
\end{equation}
where the order parameter $\varphi({\bf x})$ is an
$n$--component vector with components $\varphi_i({\bf x})$, depending on the
coordinate ${\bf x}$, $T$ is the temperature, and $k_B$ is the Boltzmann constant.
It is assumed that there exists the upper cut-off parameter $\Lambda$ (a positive finite number)
for the Fourier components of the order-parameter field $\varphi_i({\bf x})$. 
Namely, the Fourier--transformed Hamiltonian reads 
\begin{equation} \label{eq:Hf}
\frac{\cal{H}}{k_B T}= \sum\limits_{i,{\bf k}} \left( r_0+c \,{\bf k}^2 \right)
{\mid \varphi_{i,{\bf k}} \mid}^2 + uV^{-1}
\sum\limits_{i,j,{\bf k}_1,{\bf k}_2,{\bf k}_3}
\varphi_{i,{\bf k}_1} \varphi_{i,{\bf k}_2} \varphi_{j,{\bf k}_3}
\varphi_{j,-{\bf k}_1-{\bf k}_2-{\bf k}_3} \;,
\end{equation}
where $\varphi_{i,{\bf k}}=V^{-1/2} \int \varphi_i({\bf x}) \exp(-i {\bf kx}) \, d{\bf x}$
and $\varphi_i({\bf x}) = V^{-1/2} \sum\limits_{k<\Lambda} \varphi_{i,{\bf k}} \exp(i {\bf kx})$.
Moreover, the only allowed configurations of $\varphi_i({\bf x})$ are those, for which
$\varphi_{i,{\bf k}}=0$ holds at $k \equiv \mid {\bf k} \mid > \Lambda$ (therefore 
we set $\varphi_{i,{\bf k}}=0$ at $k > \Lambda$ in~(\ref{eq:Hf})). This is the limiting case $m \to \infty$ of the
model where all configurations are allowed, but Hamiltonian~(\ref{eq:Hf})
is completed by the term $\sum_{i,{\bf k}} \left( k/\Lambda \right)^{2m}
\mid \varphi_{i,{\bf k}} \mid^2$. 

We define the temperature--dependence of the Hamiltonian parameters in vicinity of 
the critical temperature $T_c$ by a linear relation
\begin{equation}
 r_0 = r_{0c} + a (T-T_c) \;,
\label{eq:r0}
 \end{equation}
where $r_{0c}$ is the critical value of $r_0$ and $a$ is a constant.
The parameters $c$ and $u$ are assumed to be $T$--independent.
For simplicity, we will consider only the case $T>T_c$ (or $r_0>r_{0c}$).

Using~(\ref{eq:H}) or~(\ref{eq:Hf}), we can easily calculate the derivative
\begin{equation}
\frac{\partial}{\partial r_0} \left( \frac{F}{k_B T} \right) = - \frac{\partial \ln Z}{\partial r_0} 
= V \left\langle  \varphi^2({\bf x}) \right\rangle = n \sum\limits_{k<\Lambda} G({\bf k}) \;,
\label{der}
\end{equation}
where $F = -k_B T \ln Z$ is the free energy, $Z= \int \exp[-H/(k_BT)] \mathcal{D} \varphi$ is the partition function
and $G({\bf k}) = \langle \mid \varphi_{i,{\bf k}} \mid^2 \rangle$ (for any $i=1,2, \ldots, n$) 
is the Fourier--transformed two--point correlation function.
In the thermodynamic limit at $T>T_c$, the sum over ${\bf k}$ in~(\ref{der}) is replaced by the integral 
according to the well known rule $\sum_{\bf k} \to  V (2 \pi)^{-d} \int d {\bf k}$
(the term with ${\bf k = 0}$ has to be separated at $T<T_c$), where $d$ is the spatial dimensionality.
The internal energy $U = - T^2 \left( \partial(F/T) / \partial T \right)_V$, calculated from~(\ref{eq:r0})
and~(\ref{der}), therefore is 
\begin{equation}
 U = -a k_B T^2 n V (2 \pi)^{-d} \int\limits_{k<\Lambda} G({\bf k}) d {\bf k} \;.
 \label{eq:Using}
\end{equation}

Consider now the singularity of $U$ and the related singularity of specific
heat $C_V$ in vicinity of the critical point at $t \to 0$, where $t = (T-T_c)/T_c$ is the
reduced temperature. We assume that the singular part of $C_V$ has the form
$\propto (\ln t)^s t^{-\alpha}$ at $t \to 0$. According to the thermodynamic relation
$C_V = (\partial U / \partial T)_V$, the corresponding singular part of $U$ is 
$\propto (\ln t)^s t^{1-\alpha}$ at $t \to 0$. Further on, we will consider the normalized quantities
$U/V$ and $C_V/V$ and represent the singularities in terms of the correlation length $\xi$,
assuming the power--law scaling $\xi \propto t^{-\nu}$ at $t \to 0$. The latter is known
to be true for the $\varphi^4$ model in three dimensions at any $n \ge 1$, as well as at $d=2$ and $n=1$.
The above relations imply $C_V^{sing} \propto \xi^{1/\nu} U^{sing}$, where  $U^{sing}$
and $C_V^{sing}$ are the leading singular parts of $U/V$ and $C_V/V$, represented in
powers of $\xi$ and $\ln \xi$ at $\xi \to \infty$. Using~(\ref{eq:Using}), it yields
\begin{equation}
C_V^{sing} = B \xi^{1/\nu} \, \left( \int_{k<\Lambda} [G({\bf k})- G^*({\bf k})] d {\bf k} \right)^{sing} \;,
\label{eq:CVsing}
\end{equation}
where $G^*({\bf k})$ is the value of $G({\bf k})$ at the critical point and $B$ is a nonzero constant. 
The superscript ``$sing$'' implies the leading singular contribution in terms of $\xi$.
Since the singular part does not include a constant contribution, it is subtracted in 
brackets of~(\ref{eq:CVsing}).

Let us denote by $C_V^{sing}(\Lambda')$ the contribution of the integration region $0<k<\Lambda'$ to~(\ref{eq:CVsing}),
where $0 < \Lambda' \le \Lambda$. Note that 
$G({\bf k})$ and $G^*({\bf k})$ always correspond to the true upper cut-off $\Lambda$.
Based on the idea that the short--wavelength contribution is irrelevant,
it has been assumed in~\cite{K2012} that $C_V^{sing}(\Lambda')$ is independent of $\Lambda'$. To the contrary,
here we allow that the amplitude of the leading singularity depends on $\Lambda'$. Namely, it is assumed
that $C_V^{sing}(\Lambda') = A(\Lambda') \, (\ln \xi)^{\lambda} \xi^{\alpha/\nu}$ holds
with $\lambda=0$ corresponding to the usual power--law scaling. In addition, we assume
that $\lim_{\Lambda' \to 0} A(\Lambda') \ne 0$ holds, implying that the long--wavelength (small $k$) contribution to
the integral in~(\ref{eq:CVsing}) is relevant. Note that the amplitude $A(\Lambda')$ is determined, considering
the limit $\xi \to \infty$ at a fixed $\Lambda'$. It means that, even at $\Lambda' \to 0$, the limit $\xi \to \infty$
is considered first and, therefore, the relevant region of small wave vectors $k \sim 1/\xi$ is always included. 
The above mentioned assumptions have been tested numerically in Sec.~\ref{sec:t}, clearly showing that they hold
in the 2D model with $\Lambda'$--dependent amplitude $A(\Lambda')$.

Since we consider the limit $\Lambda' \to 0$, it is naturally to use the scaling hypothesis 
for the correlation function, which is valid for small $k$ and large $\xi$. Namely, we have
\begin{equation}
G({\bf k}) = \sum\limits_{i \ge 0} \xi^{(\gamma - \theta_i)/\nu} g_i(k \xi) \;,
\label{eq:sc1}
\end{equation}
where $g_i(k \xi)$ are continuous scaling functions, which are finite for $0 \le k \xi < \infty$. Here $\theta_0=0$ holds and the term with $i=0$ 
describes the leading singularity, whereas the terms with $i \ge 1$ represent other 
contributions with
correction exponents $\theta_i>0$. 
The critical correlation function
\begin{equation}
G^*({\bf k}) = \sum\limits_{i \ge 0} b_i k^{(-\gamma + \theta_i)/\nu}
\label{eq:sc2}
\end{equation}
is obtained at $\xi \to \infty$, so that there exists a finite limit
\begin{equation}
\lim\limits_{z \to \infty} z^{(\gamma - \theta_i)/\nu} g_i(z) = b_i \;,
\end{equation}
where $b_i$ are constant coefficients. We allow that some of these coefficients are zero.  
Since we consider only the leading singularity of $C_V$ and the small--$k$ contribution, it is also naturally to assume
that only a finite number of correction terms is relevant in our calculations.
The assumed validity of~(\ref{eq:sc1}) and~(\ref{eq:sc2}) implies that the 
values of the exponents ensure the convergence of the integral~(\ref{eq:CVsing}) at zero lower
integration limit. It means that $d- \gamma/\nu > 0$ must hold at $\theta_i \ge 0$.

Based on the discussed here scaling assumptions, we have obtained an important and challenging result
for correction--to--scaling exponents by proving the following theorem.

\vspace*{1ex}

\textbf{Theorem.} \hspace{0ex} 
\textit{
If the leading singular part of specific heat $C_V^{sing}$~(\ref{eq:CVsing})  
has the form  $C_V^{sing} \propto (\ln \xi)^{\lambda} \xi^{\alpha/\nu}$ 
and the contribution of the region $k < \Lambda'$
has the form $C_V^{sing}(\Lambda') = A(\Lambda') \, (\ln \xi)^{\lambda} \xi^{\alpha/\nu}$ with 
$\lim_{\Lambda' \to 0} A(\Lambda') \ne 0$, if correct result in the $\lim_{\Lambda' \to 0} \lim_{\xi \to \infty}$ limit 
(considering $\xi \to \infty$ at a fixed $\Lambda'$ first)
is obtained using 
~(\ref{eq:sc1}) --(\ref{eq:sc2}) (at the conditions of validity $d-\gamma/\nu >0$ and $\theta_i \ge 0$,
$g_i(z)$ being continuous and finite for $0 \le z < \infty$ and $\lim_{z \to \infty} z^{(\gamma - \theta_i)/\nu} g_i(z)$
being finite) with a large enough finite number of terms included,
and if $\gamma + 1 -\alpha - d \nu >0$ holds, then 
\begin{enumerate}
 \item $\lim_{\Lambda' \to 0} \mid A(\Lambda') \mid \ne \infty$;
 \item the two--point correlation function 
contains a correction--to--scaling term with exponent 
\begin{equation}
\theta_{\ell} = \gamma + 1 -\alpha - d \nu \;,
\label{eq:sakariba}
\end{equation}
corresponding to a certain term with $i=\ell \ge 1$ in~(\ref{eq:sc1}).
\end{enumerate}
} 

\vspace*{1ex}

 \textbf{Proof.} \hspace*{1ex} 
Since the correlation function in~(\ref{eq:sc1}) -- (\ref{eq:sc2}) is isotropic,
$C_V^{sing}(\Lambda')$ can be written as
\begin{equation}
 C_V^{sing}(\Lambda') = B \, S(d) \, \xi^{1/\nu} \left( \int\limits_0^{\Lambda'} 
 \sum\limits_{i \ge 0} \left[ \xi^{(\gamma - \theta_i)/\nu} g_i(k \xi)
 - b_i k^{(-\gamma + \theta_i)/\nu}  \right] k^{d-1} dk \right)^{sing} \;,
\end{equation}
where $S(d)=2 \pi^{d/2} / \Gamma(d/2)$ is the surface of unit sphere in $d$ dimensions.
For any finite number of summation terms included, the integration and summation can be
exchanged, since the integral exists and converges for each of the terms separately,
according to the conditions of validity and properties of scaling functions, mentioned in 
the theorem, and the fact that $\Lambda'$ is finite.
Then, changing the integration variable to $y=k \xi$, we obtain
\begin{equation}
C_V^{sing}(\Lambda') = B \, S(d) \, \left( \sum\limits_{i \ge 0} 
\xi^{-d+(1+\gamma-\theta_i)/\nu} F_i(\Lambda' \xi) \right)^{sing} \;,
\label{eq:CVs}
\end{equation}
where
\begin{equation}
F_i(z) = \int\limits_0^z y^{d-1} \widetilde{g}_i(y) dy \qquad \mbox{with} \quad 
 \widetilde{g}_i(y) = g_i(y) - b_i y^{(-\gamma + \theta_{\ell})/\nu} \;.
\label{eq:integ}
\end{equation}

First we will prove that only one term in~(\ref{eq:CVs}) gives the leading singular contribution
in the limit $\lim_{\Lambda' \to 0} \lim_{\xi \to \infty}$. Since $C_V^{sing}(\Lambda') \propto (\ln \xi)^{\lambda} \xi^{\alpha/\nu}$
holds, only those terms can give the leading singularity at $\xi \to \infty$, which are proportional to 
$(\ln \xi)^{\lambda} \xi^{\alpha/\nu}$ in this limit. It implies that
\begin{equation}
F_i(\Lambda' \xi) \propto \left[ \ln(\Lambda' \xi) \right]^{\lambda} (\Lambda' \xi)^{\mu_i}
\label{F_ell}
\end{equation}
must hold for these terms at $\Lambda' \xi \to \infty$ with 
\begin{equation}
-d + (1+\gamma-\theta_i)/\nu + \mu_i = \alpha/\nu \;, \quad i \in \Omega \;.
\label{mu}
\end{equation}
Here $\Omega$ is the subset of indices $i$, labeling these terms.
According to the conditions of the theorem, $\Omega$ contains a finite number of indices.
If there exist several terms with $i \in \Omega$, then they all have different exponents $\mu_i$ because 
$\theta_i$ in~(\ref{mu})
are different by definition. In the limit $\lim_{\Lambda' \to 0} \lim_{\xi \to \infty}$, these
terms give contributions $\propto (\Lambda')^{\mu_i}  \, (\ln \xi)^{\lambda} \xi^{\alpha/\nu}$,
as consistent with~(\ref{eq:CVs}) and (\ref{F_ell}) -- (\ref{mu}).
Consequently, at $\Lambda' \to 0$, the amplitude is
\begin{equation}
A(\Lambda') \propto (\Lambda')^{\mu_{\ell}} \;,
\label{A}
\end{equation}
where $\mu_{\ell} = \min\limits_{i \in \Omega} \mu_i$. Thus, we 
have proven the statement that only one of the terms in~(\ref{eq:CVs})
with certain index $i=\ell$ gives the leading singularity at $\lim_{\Lambda' \to 0} \lim_{\xi \to \infty}$. 
This is not necessarily the leading term with $\ell=0$, since the integration over $k$ can
give a vanishing result due to the cancellation of positive and negative contributions. 
Formally, there is also a possibility that some terms give analytic contributions,
which are constant or proportional to an integer power of $t$ (integer power of $\xi^{-1/\nu}$).
By definition, such terms are considered as non-singular and not contributing to $C_V^{sing}(\Lambda')$.

In the following we will prove the statement $\lim_{\Lambda' \to 0} \mid A(\Lambda') \mid \ne \infty$
by assuming the opposite and deriving a contradiction. Thus, let us assume that $A(\Lambda')$ diverges
at $\Lambda' \to 0$. According to~(\ref{A}), it is possible only for $\mu_{\ell} <0$. 
Hence, from~(\ref{eq:integ}) and~(\ref{F_ell}) we find that
\begin{equation}
 F_{\ell}(z) = \int\limits_0^z y^{d-1} \widetilde{g}_{\ell}(y) dy = c_{\ell} \, (\ln z)^{\lambda} z^{\mu_{\ell}}
\label{Fellz}
 \end{equation}
holds at $\mu_{\ell} <0$ for large $z=\Lambda' \xi \to \infty$, corresponding to the 
considered here limit $\lim_{\Lambda' \to 0} \lim_{\xi \to \infty}$. Here $c_{\ell}$ is a nonzero constant,
and~(\ref{Fellz}) holds asymptotically with relative error tending to zero at $z \to \infty$.
The derivation with respect to $z$ in~(\ref{Fellz}) yields
\begin{equation}
 \widetilde{g}_{\ell}(z) = c_{\ell} \, \left[ \lambda (\ln y)^{-1} + \mu_{\ell} \right] 
 (\ln z)^{\lambda} z^{\mu_{\ell}-d}
\qquad \mbox{at} \quad z \to \infty \;.
 \end{equation}
Consequently, the integrand function with $i=\ell$ in~(\ref{eq:integ}), i.~e., $f(y)=y^{d-1} \widetilde{g}_{\ell}(y)$, converges to
 $f_{as}(y)$ at $y \to \infty$ in such a way that $(f(y)-f_{as}(y))/f_{as}(y) \to 0$, where
\begin{equation}
 f_{as}(y) = c_{\ell} \left[ \lambda (\ln y)^{-1} + \mu_{\ell} \right] (\ln y)^{\lambda} y^{\mu_{\ell}-1}
\label{fas}
 \end{equation}
is the asymptotic form of $f(y)$. It implies that, for any given finite $\varepsilon >0$, there exists
a finite $y_0>0$, such that $\mid f(y) - f_{as}(y) \mid / \mid f_{as}(y) \mid < \varepsilon$ holds
for $y> y_0$. Since $\mid f(y) \mid - \mid f_{as}(y) \mid \le \mid f(y) - f_{as}(y) \mid$ always holds,
we have also
\begin{equation}
 \frac{\mid f(y) \mid - \mid f_{as}(y) \mid}{\mid f_{as}(y) \mid} 
 < \varepsilon \qquad \mbox{for} \quad y>y_0 \;.
\label{estim}
 \end{equation}
 
At this condition, the integral~(\ref{eq:integ}) with $i= \ell$ converges at $z \to \infty$. 
To prove this statement,
the integral at $z \to \infty$ is written as $\int_0^{\infty} f(y)dy = \int_0^{y_0} f(y) dy + \int_{y_0}^{\infty} f(y) dy$.
The first integral $\int_0^{y_0} f(y) dy$ exists and it has a finite value because the scaling function $g_{\ell}(y)$ is continuous and finite
within $0 \le y  \le y_0$,
as well as $d-\gamma/\nu>0$ and $\theta_{\ell} \ge 0$ hold for the exponents.
Using~(\ref{estim}), the second integral
can be evaluated as 
\begin{equation}
\left| \int\limits_{y_0}^{\infty} f(y) dy \right| \le \int\limits_{y_0}^{\infty} \mid f(y) \mid dy
< \int\limits_{y_0}^{\infty} \mid f_{as}(y) \mid (1 + \varepsilon ) dy \;.
\label{eval}
\end{equation}
The latter integral in~(\ref{eval}) converges according to~(\ref{fas}), since $\mu_{\ell}<0$ holds.
Consequently, the integral $\lim_{z \to \infty} F_{\ell}(z) = \int_0^{\infty} f(y)dy$ also converges. 
It means that $F_{\ell}(z)$ tends to a constant at $z \to \infty$ and, according to~(\ref{eq:CVs}),
 the amplitude $A(\Lambda')$ is constant at $\Lambda' \to 0$. It contradicts the initial assumption
 that $A(\Lambda')$ diverges at $\Lambda' \to 0$, so that this assumption is false, i.~e., 
 $\lim_{\Lambda' \to 0} \mid A(\Lambda') \mid \ne \infty$.
 
Finally, we will prove the relation~(\ref{eq:sakariba}). Since  $\lim_{\Lambda' \to 0} A(\Lambda') \ne 0$
holds according to the conditions of the theorem, we have $\mu_{\ell} \le 0$ in~(\ref{A}). On the other
hand, since $\lim_{\Lambda' \to 0} \mid A(\Lambda') \mid \ne \infty$, we have $\mu_{\ell} \ge 0$.
Consequently, $\mu_{\ell}=0$ holds. Eq.~(\ref{mu}) with $i= \ell \in \Omega$ then leads to~(\ref{eq:sakariba}).
The condition $\gamma + 1 -\alpha - d \nu >0$ of the theorem implies that~(\ref{eq:sakariba})
is satisfied with $\theta_{\ell} >0$, which corresponds to a correction term with $i=\ell >0$ in~(\ref{eq:sc1})
(the term with $i=0$ gives no contribution to $C_V^{sing}(\Lambda')$ at $\lim_{\Lambda' \to 0} \lim_{\xi \to \infty}$). 
$\Box$

\vspace*{2ex}

One has to note that, according to the self--consistent scaling theory of logarithmic correction
exponents in~\cite{KJJ06}, logarithmic corrections can generally appear in $\xi$ as function of $t$, as well as in $G({\bf k})$. 
Nevertheless, our consideration 
covers the usual case of $\lambda=0$, where no logarithmic corrections are present, as well as the important
particular case of $\alpha = 0$ and $\lambda = 1$, where the logarithmic correction appears only in specific heat~\cite{KJJ06}. 
The considered here scaling forms appear to be general enough for our analysis 
of the $\varphi^4$ model below the upper critical dimension $d<4$, where $\xi$ and $G({\bf k})$ have no logarithmic corrections
according to the known results, except only for the case of the Kosterlitz--Thouless phase transition at $n=2$ and $d=2$.
According to the current knowledge about the critical phenomena, the used here scaling forms, as well as the assumed
relations for the exponents $d- \gamma/\nu >0$ and
$\gamma + 1 -\alpha - d \nu >0$ hold for $d=3$, $n \ge 1$ and also for $d=2$, $n=1$.
The other conditions of the theorem are satisfied in these cases, 
according to the provided here general arguments and numerical tests in Sec.~\ref{sec:t}.

The existence of a correction with exponent $\theta_{\ell} =3/4$ in the scalar ($n=1$) 2D $\varphi^4$ model
follows from this theorem, if $\gamma = 7/4$ and $\nu =1$ hold here, as in the 2D Ising model.
It corresponds to a correction exponent $\omega_{\ell} = \theta_{\ell}/\nu = 3/4$ in the critical 
two--point correlation function, as well as in the finite--size scaling.
Since this exponent not necessarily describes the leading correction term, the prediction is $\omega \le 3/4$
for the leading correction--to--scaling exponent $\omega$.
 An evidence for a nontrivial correction with non--integer exponent (which might be, e.~g., $1/4$)
 in the finite--size scaling of the critical real--space two--point correlation 
function of the 2D Ising model has been provided in~\cite{Kaupuzs06}, based on an exact enumeration by a transfer matrix algorithm. 
This correction, however, has a very small amplitude
and is hardly detectable. Moreover, such a correction has not been detected in susceptibility.
Usually, the scaling in the 2D Ising model is representable by trivial, i.~e., integer, correction--to--scaling exponents
when analytical background terms or ``short--distance'' terms (e.~g., a constant contribution to susceptibility) are separated
-- see, e.~g.,~\cite{YP02,CGNP11,CHP02} and references therein. The discussions have been focused on the existence 
of irrelevant variables~\cite{Sokal,Perk}. In particular, the high--precision calculations
in~\cite{Perk} have shown that the conjecture by Aharony and Fisher
about the absence of such variables~\cite{AF80,AF83} fails.

The above mentioned theorem predicts the existence of nontrivial correction--to--scaling exponents in the
2D $\varphi^4$ model. 
It can be expected that the nontrivial correction terms of the $\varphi^4$ model 
usually do not show up or cancel in the 2D Ising model. This idea is not new. Based on the standard field--theoretical
treatments of the $\varphi^4$ model, $\omega = 4/3$
has been conjectured for the leading nontrivial scaling corrections at $n=1$ and $d=2$ in~\cite{Justin,BF84}. However, it contradicts our theorem,
which yields $\omega \le 3/4$. This discrepancy is interpreted as a failure of the standard
perturbative methods --- see~\cite{K2012x} and the discussions in~\cite{K2012}.
One has to note that the alternative perturbative approach of~\cite{K_Ann01}, predicting 
$\omega_{\ell} = \ell \eta$ (where $\ell \ge 1$ is an integer) with $\eta = 2 - \gamma/\nu = 1/4$ for $n=1$ and $d=2$, is consistent with
this theorem.

\section{Monte Carlo simulation of the lattice $\varphi^4$ model}
\label{sec:MCs}

We have performed MC simulations of the scalar 2D $\varphi^4$ model on square lattice
with periodic boundary conditions. The Hamiltonian $\mathcal{H}$ is given by
\label{sec:model}
\begin{equation}
\frac{\mathcal{H}}{k_B T}  = - \beta \sum\limits_{\langle i j \rangle}
\varphi_i \varphi_j + \sum\limits_i \left( \varphi_i^2 + \lambda \left( \varphi_i^2 -1 \right)^2 \right) \;,
\end{equation}
where $-\infty < \varphi_i < \infty$ is a continuous scalar order parameter at the $i$-th lattice site, and $\langle ij \rangle$
denotes the set of all nearest neighbors. This notation is related to the one of~\cite{Hasenbusch} via
$\beta= 2 \kappa$ and $\varphi = \phi$. We have denoted the coupling constant at $\varphi_i \varphi_j$ by $\beta$ to outline
the similarity with the Ising model.

Swendsen-Wang and Wolff cluster algorithms are known to be very efficient for MC simulations of the Ising model 
in vicinity of the critical point~\cite{MC}.
However, these algorithms update only the spin orientation, and therefore are not ergodic
for the $\varphi^4$ model. The problem is solved using the hybrid algorithm, where a cluster algorithm is
combined with Metropolis sweeps. This method has been applied  to the 
3D $\varphi^4$ model in~\cite{Hasenbusch}. In our simulations, we have applied one Metropolis sweep after
each $N_W$ Wolff single cluster algorithm steps. Following~\cite{Hasenbusch}, a new value of the order parameter 
is chosen as $\varphi'_i = \varphi_i + s (r - 1/2)$ in one Metropolis step (this value being either accepted or 
rejected, as usually) where $s$ is a constant and $r$ is a random number
from a set of uniformly distributed random numbers within $[0, 1]$. 
Here $N_W$ and $s$ are considered as optimization parameters, allowing to reach
the smallest statistical error in a given simulation time. 
We have chosen $N_W$ such that $N_W \langle c \rangle /L^2$
is about $2/3$ or $0.6$, where $\langle c \rangle$ is the mean cluster size. The optimal choice of $s$ depends on the
Hamiltonian parameters. Our simulations have been performed at $\lambda = 0.1$, $\lambda = 1$, $\lambda = 10$
and at such values of $\beta$, which correspond to $U=\langle m^4 \rangle / \langle m^2 \rangle^2 = 1.1679229 \approx U^*$
and $U=2$, $m$ being the magnetization per spin. 
Here $U^*$ is the $\lambda$--independent (universal) critical value of $U$, which has been evaluated 
as $U^*=1.1679229 \pm 0.0000047$ in~\cite{Sokal}. 
At $U = 1.1679229$, we have chosen $s=4$ for $\lambda=0.1$, $s=4$ for $\lambda=1$ and
$s=3$ for $\lambda = 10$. At $U = 2$,
the corresponding values 
are $s=3.5$, $s=3$ and $s=2$. 
For comparison, $s=3$ has been used in~\cite{Hasenbusch}.

We have used the iterative method of~\cite{KMR_2010} to find $\beta$,
corresponding to certain value of $U$, as well as a set of statistical averaged quantities at this $\beta$,
called the pseudo-critical coupling $\widetilde{\beta}_c(L)$.
We have performed high statistics simulations for evaluation of the derivative $\partial U / \partial \beta$ and 
the susceptibility $\chi = N \langle m^2 \rangle$, where $N=L^2$ is the total number of spins.
According to the Boltzmann statistics, the derivative with respect to $\beta$ for any quantity $\langle \mathcal{A} \rangle$ is calculated as
\begin{equation}
\frac{\partial}{\partial \beta} \langle \mathcal{A} \rangle = N \left[ \langle \mathcal{A} \rangle \langle \varepsilon \rangle
- \langle \mathcal{A} \varepsilon \rangle \right] \;,
\label{eq:atvas}
\end{equation}
where $\varepsilon = -N^{-1} \sum_{\langle ij \rangle} \varphi_i \varphi_j$.

For each lattice size $L$, the quantities $\chi$ and $\partial U / \partial \beta$ have been estimated from $100$ iterations (simulation bins) 
in vicinity of $\beta = \widetilde{\beta}_c(L)$, collected from
one or several simulation runs, discarding first $10$ iterations of each run for equilibration.
One iteration included $10^6$ steps of the hybrid algorithm, each consisting of one Metropolis sweep and $N_W$
Wolff algorithm steps, as explained before. To test the accuracy of our iterative method, we have performed
some simulations (for $U=2$ and $\lambda=0.1$) with $2.5 \times 10^{5}$ hybrid algorithm steps in one iteration, and have verified that
the results well agree with those for $10^6$ steps. Moreover, we have used two different
pseudo-random number generators, the same ones as in~\cite{KMR_2011}, to verify that the results agree within the 
statistical error bars.

Note that the quantity $U$ is related to the Binder cumulant $B=1-U/3$~\cite{MHB86}. In the thermodynamic limit, we have $B=0$ ($U=3$) 
above the critical point, i.~e., at $T>T_c$ or $\beta < \beta_c$, and $B=2/3$ ($U=1$) at $T<T_c$ or $\beta>\beta_c$. Thus, 
the pseudo-critical coupling $\widetilde{\beta}_c(L)$, corresponding
to a given $U$ in the range of $1<U<3$, tends to the true critical coupling $\beta_c$ at $L \to \infty$. 
The iterative algorithm of~\cite{KMR_2010} is valid for any $1<U<3$. However, it can be useful
to choose $U \approx U^*$ for MC analysis. In particular, it allows us to obtain $\widetilde{\beta}_c(L)$ values closer to the
true critical coupling $\beta_c$. On the other hand, it is crucial for our MC analysis in Sec.~\ref{sec:MCanal} to have the 
data for at least two remarkably different values of $U$. Therefore, 
we have chosen one $U$ value, $U=2$, in the middle of the interval $1<U<3$ and the other one, $U=1.1679229$, close
to the critical value $U^*$ at $\beta=\beta_c$ and $L \to \infty$. 

MC simulations have been performed for lattice sizes $4 \le L \le L_{\mathrm{max}}$ whith $L_{\mathrm{max}} = 256$
at $\lambda=10$, $L_{\mathrm{max}} = 384$ at $\lambda=1$ and $L_{\mathrm{max}} = 1536$ at $\lambda=0.1$.
This choice of $L_{\mathrm{max}}$ is motivated by the fact that at $\lambda=0.1$ we have observed
an interesting scaling behavior and, therefore, the simulations have been extended up to $L=1536$ for a refined analysis.  

In addition, we have performed some simulations with the hybrid algorithm at certain fixed values of the reduced temperature
$t = 1-\beta/\beta_c$ and have evaluated the Fourier--transformed two--point correlation function $G({\bf k})$ 
and its derivative $\partial G({\bf k})/ \partial \beta$ in order to test the conditions of the theorem
in Sec.~\ref{sec:analytical}. These results are discussed in Sec.~\ref{sec:t}.

A parallel algorithm, similar to that one used in~\cite{KMR_2010}, helped us to speed up the simulations. 
The Wolff algorithm has been parallelized in this way, whereas the usual
ideas of splitting the lattice in slices~\cite{MC} have been applied to parallelize the Metropolis algorithm. 
In the current application, the parallel code showed a quite good scalability (for Wolff, as well as Metropolis, algorithms)
up to 8 processors available on one node of the cluster. The simulation results for $\chi$ and $\partial U / \partial \beta$
are collected in Tabs.~\ref{tab1} to~\ref{tab6}.

\begin{table}
\caption{The values of $\widetilde \beta_c$, as well as $\chi/L^{7/4}$, and
$-(\partial U /\partial \beta)/L$ at $\beta = \widetilde{\beta}_c$ for $\lambda=0.1$ and $U=2$ depending on the lattice size $L$.}
\label{tab1}
\begin{center}
\begin{tabular}{|c|c|c|c|}
\hline
\rule[-2mm]{0mm}{7mm}
L & $\widetilde \beta_c$ & $\chi/L^{7/4}$  & $-(\partial U /\partial \beta)/L$  \\
\hline
4 & 0.549398(42) & 0.60791(23) & 2.4344(16)  \\
6 & 0.562326(28) & 0.50107(21) & 2.5045(17)  \\
8 & 0.570550(19) & 0.44694(19) & 2.5492(21)  \\
12  & 0.580455(14) & 0.39460(21) & 2.5900(22)  \\
16  & 0.5861408(94) & 0.36991(17) & 2.6112(26)  \\
24  & 0.5924039(62) & 0.34936(16) & 2.6459(26)  \\
32  & 0.5957584(45) & 0.34116(15) & 2.6663(30)  \\
48  & 0.5992406(34) & 0.33538(14) & 2.6881(27)  \\
64  & 0.6010332(23) & 0.33412(13) & 2.7129(31)  \\
96  & 0.6028383(15) & 0.33359(14) & 2.7273(36)  \\
128  & 0.6037470(11) & 0.33396(14) & 2.7351(40)  \\
192  & 0.60465804(69) & 0.33486(14) & 2.7592(37)  \\
256  & 0.60511333(60) & 0.33537(12) & 2.7614(38)  \\
384  & 0.60556996(40) & 0.33648(12) & 2.7745(38) \\
512  & 0.60579773(41) & 0.33701(11) & 2.7780(40) \\
768  & 0.60602518(20) & 0.337618(99) & 2.7836(37) \\
1024   & 0.60613849(13) & 0.33780(13) & 2.7827(44) \\
1536   & 0.606252278(88) & 0.33825(11) & 2.7890(40) \\
\hline
\end{tabular}
\end{center}
\end{table}

\begin{table}
\caption{The same quantities as in Tab.~\ref{tab1} for $\lambda=0.1$ and $U=1.1679229 \approx U^*$.}
\label{tab2}
\begin{center}
\begin{tabular}{|c|c|c|c|}
\hline
\rule[-2mm]{0mm}{7mm}
L & $\widetilde \beta_c$ & $\chi/L^{7/4}$  & $-(\partial U /\partial \beta)/L$  \\
\hline
4 & 0.657515(35) & 2.34779(47) & 0.80400(54)  \\
6 & 0.631043(25) & 1.92047(42) & 0.86944(59)  \\
8 & 0.620578(18) & 1.70494(34) & 0.91842(62)  \\
12 & 0.612621(12) & 1.49436(34) & 0.98840(78) \\
16 & 0.6097815(87) & 1.39473(29) & 1.03343(72) \\
24 & 0.6077860(60) & 1.30274(25) & 1.08458(88) \\
32 & 0.6071398(45) & 1.26197(23) & 1.11290(88) \\
48 & 0.6067318(29) & 1.22651(20) & 1.1405(10)  \\
64 & 0.6065998(21) & 1.21076(20) & 1.1527(10)  \\
96 & 0.6065241(13) & 1.19834(18) & 1.1662(11) \\
128 & 0.60649879(98) & 1.19250(16) & 1.1691(12) \\
192 & 0.60648734(65) & 1.18840(17) & 1.1773(12) \\
256 & 0.60648276(42) & 1.18684(16) & 1.1821(12) \\
384 & 0.60648026(37) & 1.18525(16) & 1.1827(13) \\
512 & 0.60647976(24) & 1.18469(14) & 1.1826(13) \\
768 & 0.60647922(16) & 1.18400(15) & 1.1824(13) \\
1024 & 0.60647921(13) & 1.18389(15) & 1.1822(15) \\
1536 & 0.606479145(90) & 1.18358(14) & 1.1814(15) \\
\hline
\end{tabular}
\end{center}
\end{table}

\begin{table}
\caption{The same quantities as in Tab.~\ref{tab1} for $\lambda=1$ and $U=2$.}
\label{tab3}
\begin{center}
\begin{tabular}{|c|c|c|c|}
\hline
\rule[-2mm]{0mm}{7mm}
L & $\widetilde \beta_c$ & $\chi/L^{7/4}$  & $-(\partial U /\partial \beta)/L$  \\
\hline
4 & 0.512944(44) & 0.315623(68) & 1.27395(46)  \\
6 & 0.562964(29) & 0.285305(62) & 1.26851(65)  \\
8 & 0.590002(23) & 0.270748(70) & 1.26492(60)  \\
12 & 0.618562(16) & 0.257116(64) & 1.25903(71) \\
16 & 0.633498(12) & 0.251359(59) & 1.25682(77) \\
24 & 0.6488801(88) & 0.246529(69) & 1.25454(83) \\
32 & 0.6567123(57) & 0.244758(59) & 1.25660(94) \\
48 & 0.6646251(41) & 0.243446(55) & 1.25736(90) \\
64 & 0.6686081(32) & 0.243132(52) & 1.26232(87) \\
96 & 0.6726031(19) & 0.242971(54) & 1.26114(90) \\
128 & 0.6745993(15) & 0.242899(53) & 1.2624(10) \\
192 & 0.6766057(11) & 0.243177(58) & 1.2653(12) \\
256 & 0.67760534(71) & 0.243253(50) & 1.26330(93) \\
384 & 0.67860417(52) & 0.243323(53) & 1.26412(99) \\
\hline
\end{tabular}
\end{center}
\end{table}

\begin{table}
\caption{The same quantities as in Tab.~\ref{tab1} for $\lambda=1$ and $U=1.1679229 \approx U^*$.}
\label{tab4}
\begin{center}
\begin{tabular}{|c|c|c|c|}
\hline
\rule[-2mm]{0mm}{7mm}
L & $\widetilde \beta_c$ & $\chi/L^{7/4}$  & $-(\partial U /\partial \beta)/L$  \\
\hline
4 & 0.722627(50) & 1.04218(11) & 0.42994(15) \\
6 & 0.697790(35) & 0.968521(95) & 0.46880(16) \\
8 & 0.689603(26) & 0.932407(93) & 0.48802(19) \\
12 & 0.684134(15) & 0.898059(93) & 0.50719(22) \\
16 & 0.682362(13) & 0.88219(11) & 0.51584(28) \\
24 & 0.6812691(70) & 0.868287(84) & 0.52464(27) \\
32 & 0.6809349(57) & 0.862141(79) & 0.52862(24) \\
48 & 0.6807164(37) & 0.856758(66) & 0.53201(25) \\
64 & 0.6806597(26) & 0.854718(69) & 0.53383(30) \\
96 & 0.6806247(19) & 0.852876(74) & 0.53536(33) \\
128 & 0.6806146(16) & 0.851979(80) & 0.53527(36) \\
192 & 0.68060666(96) & 0.851343(62) & 0.53597(34) \\
256 & 0.68060766(79) & 0.851228(63) & 0.53655(37) \\
384 & 0.68060481(50) & 0.850877(85) & 0.53608(44) \\
\hline
\end{tabular}
\end{center}
\end{table}

\begin{table}
\caption{The same quantities as in Tab.~\ref{tab1} for $\lambda=10$ and $U=2$.}
\label{tab5}
\begin{center}
\begin{tabular}{|c|c|c|c|}
\hline
\rule[-2mm]{0mm}{7mm}
L & $\widetilde \beta_c$ & $\chi/L^{7/4}$  & $-(\partial U /\partial \beta)/L$  \\
\hline
4 & 0.287517(24) & 0.367807(43) & 1.53677(43)  \\
8 & 0.374876(16) & 0.332360(53) & 1.35076(57)  \\
16  & 0.4217519(94) & 0.314668(54) & 1.27009(55)  \\
32  & 0.4461113(40) & 0.305725(47) & 1.23283(56)  \\
64  & 0.4585487(24) & 0.301387(50) & 1.21446(69)  \\
128 & 0.4648281(11) & 0.299070(47) & 1.20451(70)  \\
256 & 0.46798547(75) & 0.297843(55) & 1.19889(89) \\
\hline
\end{tabular}
\end{center}
\end{table}

\begin{table}
\caption{The same quantities as in Tab.~\ref{tab1} for $\lambda=10$ and $U=1.1679229 \approx U^*$.}
\label{tab6}
\begin{center}
\begin{tabular}{|c|c|c|c|}
\hline
\rule[-2mm]{0mm}{7mm}
L & $\widetilde \beta_c$ & $\chi/L^{7/4}$  & $-(\partial U /\partial \beta)/L$  \\
\hline
4 & 0.465260(24) & 1.005811(34) & 0.532729(66)  \\
8 & 0.470103(15) & 1.025510(55) & 0.51648(13)  \\
16  & 0.4709851(82) & 1.032848(67) & 0.50965(18) \\
32  & 0.4711260(32) & 1.035132(50) & 0.50721(14)  \\
64  & 0.4711546(20) & 1.036045(62) & 0.50704(20)  \\
128 & 0.4711559(11) & 1.036188(67) & 0.50673(23)  \\
256 & 0.47115644(49) & 1.036266(58) & 0.50653(21) \\
\hline
\end{tabular}
\end{center}
\end{table}

\section{Estimation of the critical coupling}
\label{sec:crp}

\begin{figure}
\begin{center}
\includegraphics[width=0.32\textwidth]{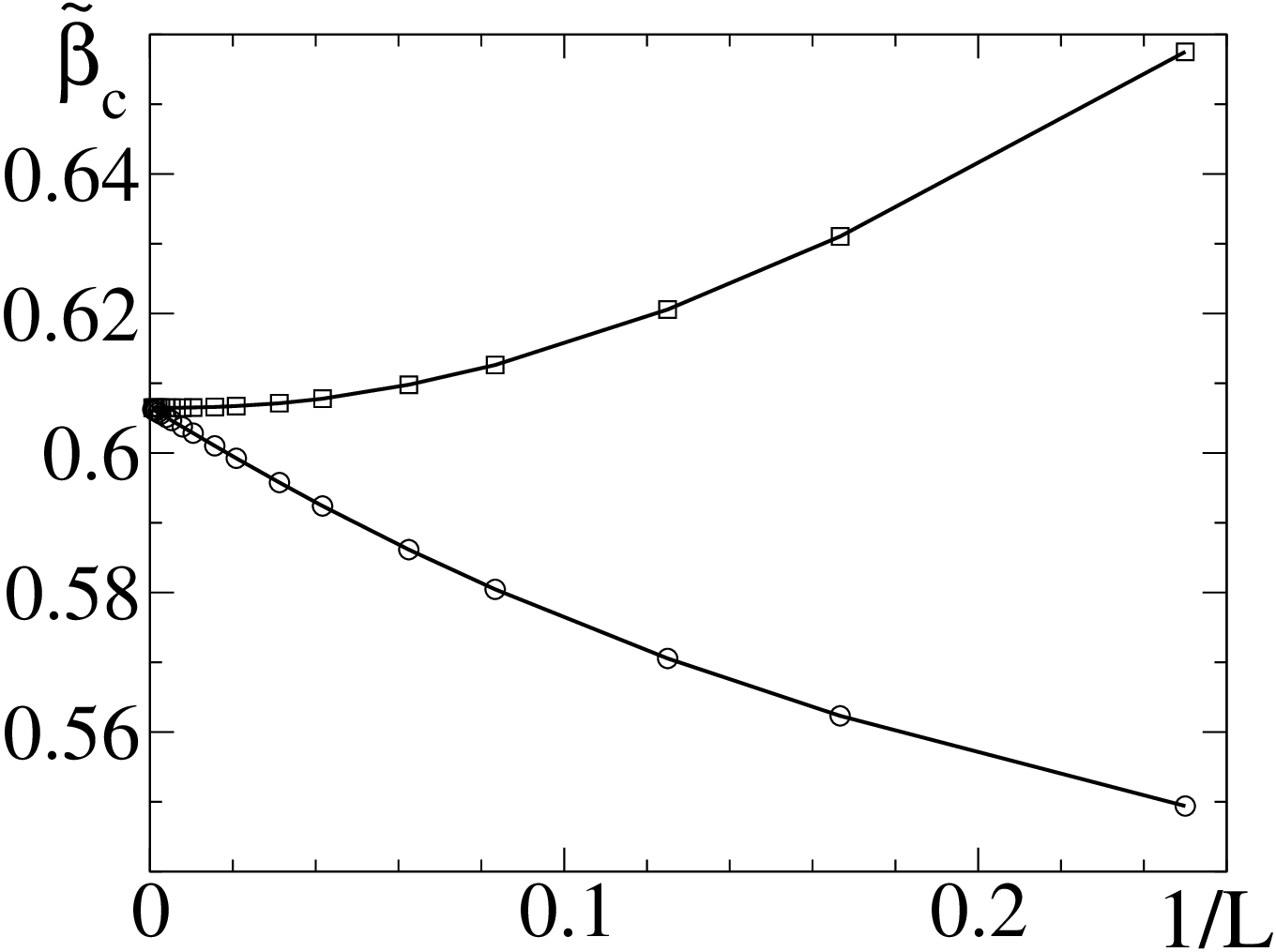} 
\hfill
\includegraphics[width=0.32\textwidth]{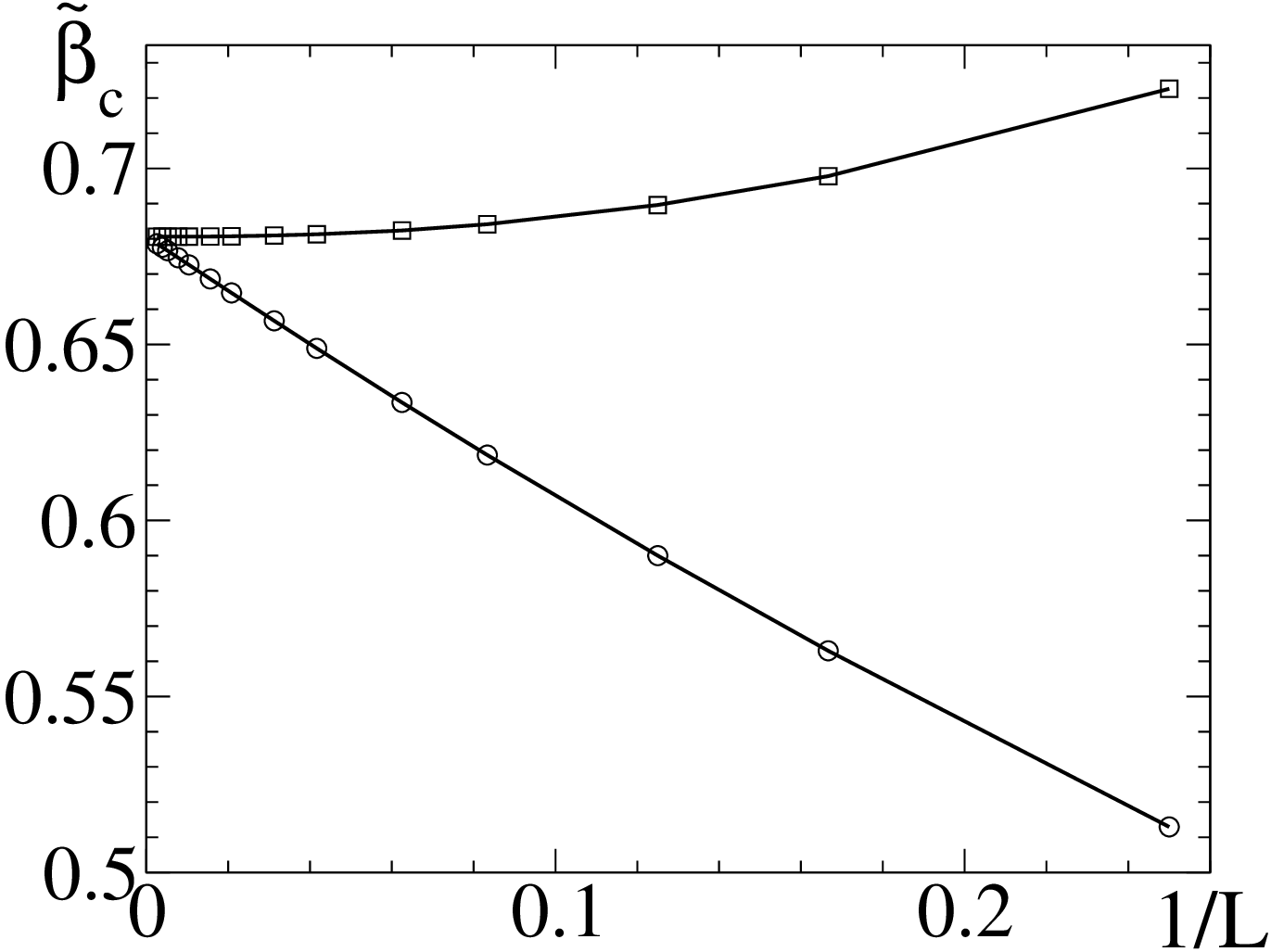} 
\hfill
\includegraphics[width=0.32\textwidth]{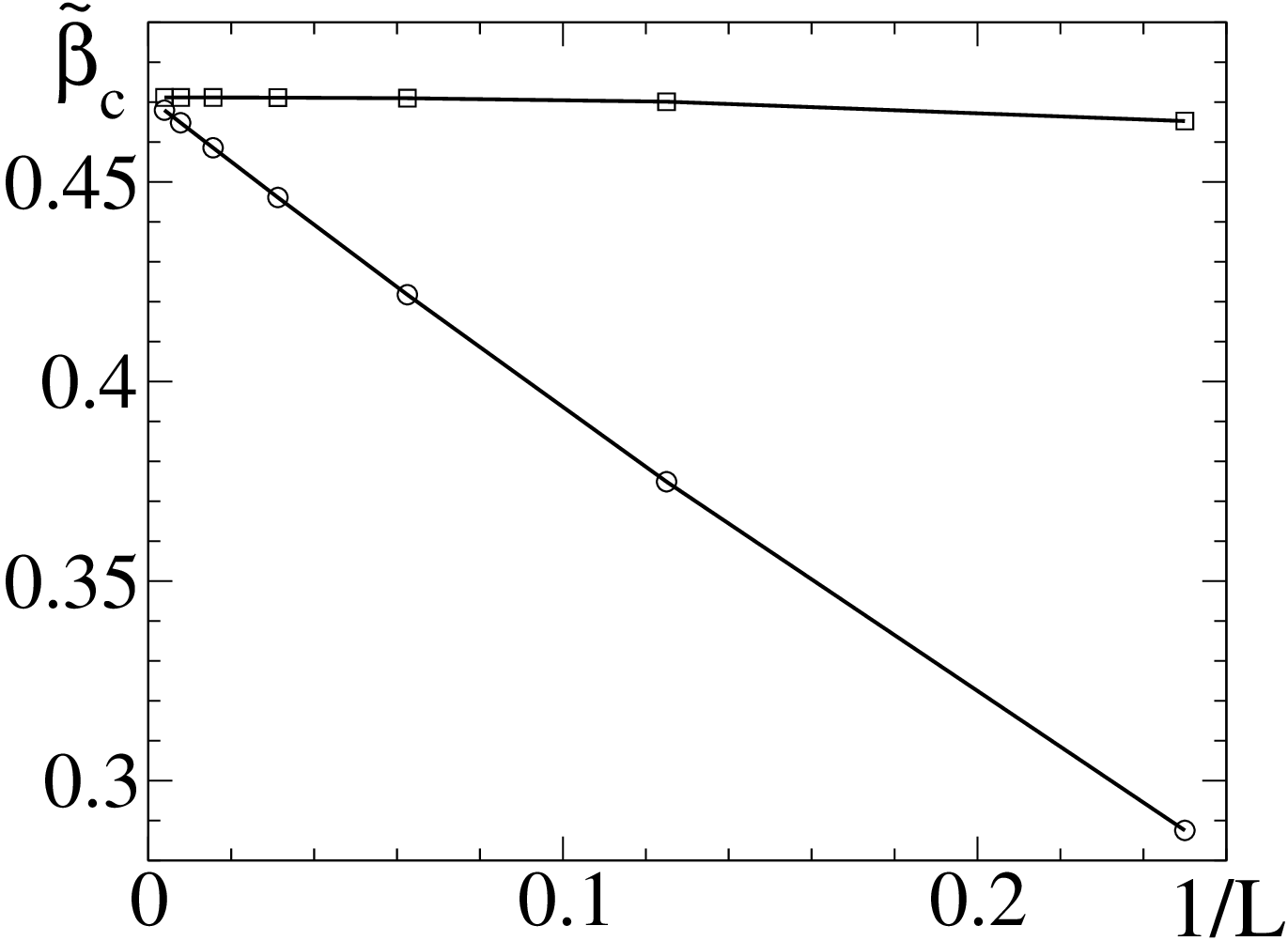}
\end{center}
\caption{The pseudo-critical coupling $\widetilde \beta_c$ vs $1/L$ for $\lambda=0.1$ (left),
$\lambda=1$ (middle) and $\lambda=10$ (right). The upper plots (squares) and the lower plots (circles) refer to the cases 
$U=1.1679229 \approx U^*$ and $U=2$, respectively. Statistical errors are much smaller than the symbol size.}
\label{betac}
\end{figure}

According to the finite--size scaling theory, $U$ behaves asymptotically as\linebreak 
$U=F \left((\beta - \beta_c)L^{1/\nu} \right)$ (see, e.~g., the references in~\cite{Hasenbusch}) for large
lattice sizes in vicinity of the critical point, where $F(z)$ is a smooth function of $z$. Hence, 
the pseudo-critical coupling $\widetilde \beta_c$ behaves as
\begin{equation}
 \widetilde \beta_c = \beta_c +  a L^{-1/\nu}
 \label{eq:betac}
\end{equation}
at large $L$, where the coefficient $a$ depends on $U$ and $\lambda$. Since $\nu=1$ holds in this model, it
is meaningful to plot $\widetilde \beta_c$ vs $1/L$ as it is done in Fig.~\ref{betac}.

At the critical $U$ value, $U=U^*$, the coefficient $a$ vanishes and the asymptotic convergence of $\widetilde \beta_c$ to the
critical coupling $\beta_c$ is faster than $\sim 1/L$. As one can judge from Fig.~\ref{betac}, it occurs at $U=1.1679229 \approx U^*$ 
for all $\lambda$. In this sense $U^*$ is universal. The estimate $U^*=1.1679229 \pm 0.0000047$ has been obtained in~\cite{Sokal}
for the 2D Ising model, corresponding to the limit $\lambda \to \infty$.

The critical coupling $\beta_c$ can be evaluated by fitting the $\widetilde \beta_c$ data at $U=2$ to the ansatz~(\ref{eq:betac}).
Alternatively, the data for $U=1.1679229 \approx U^*$ can be used. The coefficient $a$ in~(\ref{eq:betac}) vanishes at
$U=U^*$ and the convergence to $\beta_c$ is very fast in this case. Therefore, the value of  $\widetilde \beta_c(L)$ at the maximal lattice
size $L=L_{\mathrm{max}}$ for  $U=1.1679229$ can be assumed as a reasonable estimate of $\beta_c$, and
$\pm \mid \widetilde \beta_c(L_{\mathrm{max}}) - \widetilde \beta_c(L_{\mathrm{max}}/2) \mid$
can be assumed as error bars for the systematical errors. One has to take into account also
the statistical errors in $\widetilde \beta_c(L_{\mathrm{max}})$ and 
$\mid \widetilde \beta_c(L_{\mathrm{max}}) - \widetilde \beta_c(L_{\mathrm{max}}/2) \mid$. The estimates
$\beta_c = 0.60647915 \pm 0.00000035$ at $\lambda=0.1$, $\beta_c = 0.680605 \pm 0.000004$ at $\lambda = 1$
and $\beta_c = 0.4711564 \pm 0.0000020$ at $\lambda=10$ have been obtained by this method.

These estimates well agree with those obtained by fitting the data for $U=2$ to~(\ref{eq:betac}).
These fits, however, are somewhat problematic, since it is not possible to fit reasonably well more
than three data points. Since the data for relatively large lattice sizes are available at $\lambda=0.1$
and $U=2$, the problem is resolved by using a refined ansatz
\begin{equation}
 \widetilde \beta_c = \beta_c +  a_1 L^{-1/\nu} + a_2 L^{-\omega - 1/ \nu}  \;. 
 \label{eq:betacr}
\end{equation}
Here we set $\nu=1$, as in the 2D Ising model. If corrections to scaling are such as in the 2D Ising model, 
then we have $\omega=1$ in~(\ref{eq:betacr}).
However, according to the analytical arguments in Sec.~\ref{sec:analytical} and our following numerical analysis, 
smaller values of $\omega$ can be expected, such as $3/4$, $1/2$ or even $1/4$.
Fortunately, the fits within $L \in [L_{\mathrm{min}},1536]$ with $L_{\mathrm{min}}=192$ are acceptable and the fitted value of $\beta_c$ is very robust,
i.~e., it only weakly depends on $\omega$. Namely, we obtain $\beta_c = 0.60647936(24)$ with $\chi^2/\mathrm{d.o.f.}=1.18$ (where
$\chi^2/\mathrm{d.o.f.}$ is the value of $\chi^2$ per degree of freedom of the 
fit~\cite{Recipes,MC}) at $\omega=1$, $\beta_c = 0.60647915(30)$ with $\chi^2/\mathrm{d.o.f.}=1.14$ at $\omega=0.5$, and 
$\beta_c = 0.60647897(35)$ with $\chi^2/\mathrm{d.o.f.}=1.15$ at $\omega=0.25$. 
Moreover, the fits with $L_{\mathrm{min}}=256$ well confirm these results. Taking into account the statistical,
as well as the systematical errors (due to the uncertainty in $\omega$ and influence of $L_{\mathrm{min}}$),
our estimate of the critical coupling at $\lambda=0.1$ by this method is $\beta_c= 0.606479 \pm 0.000001$.

According to~(\ref{eq:H}), the fluctuations of $\varphi_i^2$ are suppressed
at $\lambda \to \infty$ in such a way that $\varphi_i^2 \to 1$ holds for relevant spin configurations with finite values
of $H/(k_BT)$ per spin. It means that the actual $\varphi^4$ model becomes equivalent to the Ising model,
where $\varphi_i = \pm 1$, in the limit $\lambda \to \infty$, further called the Ising limit.
Thus, it is not surprising that $\beta_c$ approaches the known exact value 
$\frac{1}{2} \ln \left(1+ \sqrt{2} \right) = 0.44068679 \ldots$ of the 2D Ising model~\cite{Baxter} when $\lambda$ becomes large.

It is somewhat unexpected that $\beta_c$ appears to be a non-monotonous function of $\lambda$. 
It can be explained by two competing effects. On the one hand, fluctuations increase with decreasing
of $\lambda$, and therefore $\beta_c$ tends to increase. Indeed, $\beta_c$  at $\lambda=1$ is remarkably
larger than that at $\lambda=10$. On the other hand, an effective interaction between spins becomes stronger for small 
$\lambda$ because $\langle \mid \varphi_i \mid \rangle$ and therefore also $\langle \varphi_i \varphi_j \rangle$ 
for neighboring spins increases in this case. It can explain the fact that $\beta_c$ at $\lambda = 0.1$ is
slightly smaller than that at $\lambda = 1$.

\section{Numerical test of the conditions of the theorem}
\label{sec:t}

Let us consider the quantity
\begin{equation}
 \Psi = 2 \pi \frac{\partial}{\partial \beta} \langle \varphi^2 \rangle
= \frac{2 \pi}{L^2} \sum\limits_{\bf k} \frac{\partial}{\partial \beta} G({\bf k})
\label{eq:Psi}
\end{equation}
in the scalar 2D lattice $\varphi^4$ model, where $G({\bf k}) = \langle \mid \varphi_{\bf k} \mid^2 \rangle$ with 
$\varphi_{\bf k} = L^{-1} \sum_{\bf x} \varphi({\bf x}) \exp(-i{\bf kx})$ is
the Fourier--transformed two--point correlation function, and the summation in~(\ref{eq:Psi}) takes place over 
wave vectors ${\bf k} = (k_x,k_y)$ with components $k_x = 2 \pi j/L$ and $k_y = 2 \pi l/L$,
$j$ and $l$ being integers ranging from $1-L+[L/2]$ to $[L/2]$, where $[L/2]$ denotes the integer part of $L/2$.
In the thermodynamic limit $L \to \infty$ above the critical point ($\beta < \beta_c$), the sum in~(\ref{eq:Psi})
becomes an integral 
\begin{equation}
 \Psi = \frac{1}{2 \pi} \int\limits_{\mid k_x \mid, \mid k_y \mid \le \pi} \frac{\partial}{\partial \beta} G({\bf k}) d{\bf k}
 =  -\frac{1}{2 \pi \beta_c} \, \frac{\partial}{\partial t} \int\limits_{\mid k_x \mid ,\mid k_y \mid \le \pi}  G({\bf k}) d{\bf k} \,
\label{eq:Psii}
\end{equation}
where $t=1-\beta/\beta_c$ is the reduced temperature. 
The same quantity can be considered in the continuous $\varphi^4$ model of Sec.~\ref{sec:analytical} in two dimensions
with the only difference that the integration region is $k=\sqrt{k_x^2+k_y^2}<\Lambda$ and $t$ is
defined by~(\ref{eq:r0}) there. Moreover, the small-$k$ contributions are similar in both cases,
since the correlation function is isotropic at $k \to 0$ in both models.
According standard universality arguments, these contributions thus have singularities
of the same kind, which are described by the same critical exponents and logarithmic corrections
at $t \to 0$. Moreover, according to~(\ref{eq:Using}) and $C_V = (\partial U / \partial T)_V$,
we have an equivalent to~(\ref{eq:CVsing}) representation of $C_V^{sing}$ in the form of
\begin{equation}
 C_V^{sing} \propto \left( \frac{\partial}{\partial t} \int\limits_{k<\Lambda} G({\bf k}) d {\bf k} \right)^{sing} \;,
 \label{eq:Cvs2}
\end{equation}
so that the small-$k$ contribution to specific heat in the continuous model also has the singularity of this kind.
Since $G({\bf k})$ is isotropic
at $k \to 0$, the contribution of a small-$k$ region $k<\Lambda'\ll \pi$, denoted as $\Psi(\Lambda')$, 
can be represented as
\begin{equation}
 \Psi(\Lambda') =  \int\limits_0^{\Lambda'} k \, \frac{\partial}{\partial \beta} G(k) \, dk
 = \sum\limits_{0<k<\Lambda'} k \, \frac{\partial}{\partial \beta} G(k) \, \Delta k 
\qquad \mbox{at} \quad \Delta k \to 0
 \label{eq:Psiii}
 \end{equation}
in the thermodynamic limit at $\Lambda' \to 0$, where $G(k)$ is the correlation
function in the $\langle 10 \rangle$ crystallographic direction, i.~e., at ${\bf k}=(k,0)$ or 
${\bf k}=(0,k)$, and the summation runs over $k$ values  $l \Delta k$ with integer $l > 0$
and $\Delta k = 2 \pi/L$.

 In the following we consider the quantity
\begin{equation}
 \Phi = \sum\limits_{0< k \le \pi} k \, \frac{\partial}{\partial \beta} G(k) \, \Delta k \;, 
\label{eq:Phi}
\end{equation}
which has the same small-$k$ contribution as $\Psi$, but is more convenient for simulations.
The small-$k$ contribution can be calculated from
\begin{equation}
 \Phi(\Lambda') = \Phi(\pi) - \Delta \Phi(\Lambda') \;,
\end{equation}
where $\Phi(\pi) \equiv \Phi$ and
\begin{equation}
 \Delta \Phi(\Lambda') = \sum\limits_{\Lambda' \le  k \le \pi} k \, \frac{\partial}{\partial \beta} G(k) \, \Delta k \;, 
\label{eq:DPhi}
\end{equation}
is the short-wavelength, i.~e., large-$k$ contribution.

The correlation function in the $\langle 10 \rangle$ direction is calculated as
$\langle \mid \varphi_{\bf k} \mid^2 \rangle$ for ${\bf k}=(k,0)$, i.~e.,
\begin{equation}
 G(k) = \left\langle L^{-2} \left[ \left( \sum\limits_{x=0}^{L-1} \sigma(x) \cos(kx) \right)^2 
 + \left( \sum\limits_{x=0}^{L-1} \sigma(x) \sin(kx) \right)^2 \right] \right\rangle \;,
\end{equation}
where $\sigma(x)=\sum_{y=0}^{L-1} \varphi(x,y)$ with  $\varphi(x,y)=\varphi({\bf r})$ at ${\bf r}=(x,y)$. 
The result for $\langle 01 \rangle$ direction is obtained by exchanging $x$ and $y$.
We have averaged over both equivalent cases to obtain more accurate values of $G(k)$ from MC simulations.
The derivative $\partial G(k) /\partial \beta$ is calculated using~(\ref{eq:atvas}).

\begin{figure}
\begin{center}
\includegraphics[width=0.5\textwidth]{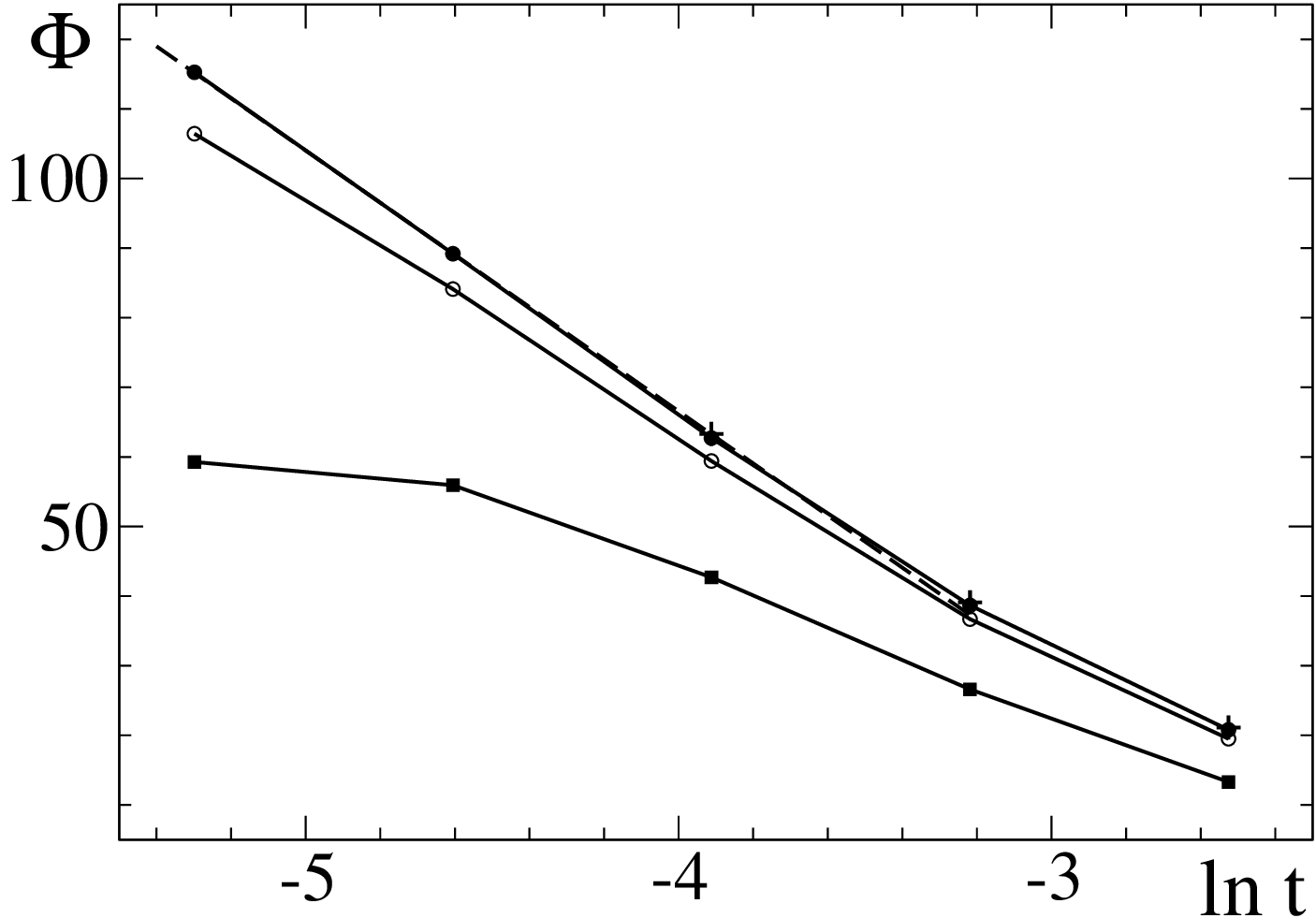}
\end{center}
\caption{The $\Phi$ vs $\ln t$ plots at $tL=1.28$ (squares), $tL=2.56$ (empty circles),
$tL=5.12$ (solid circles) and $tL=10.24$ (pluses). The dashed straight line shows that
the plot at $tL=5.12$ is almost linear within $0.005 \le t \le 0.04$.
Statistical errors are smaller than symbol size.}
\label{fig:Phi}
\end{figure}

\begin{figure}
\begin{center}
\includegraphics[width=0.5\textwidth]{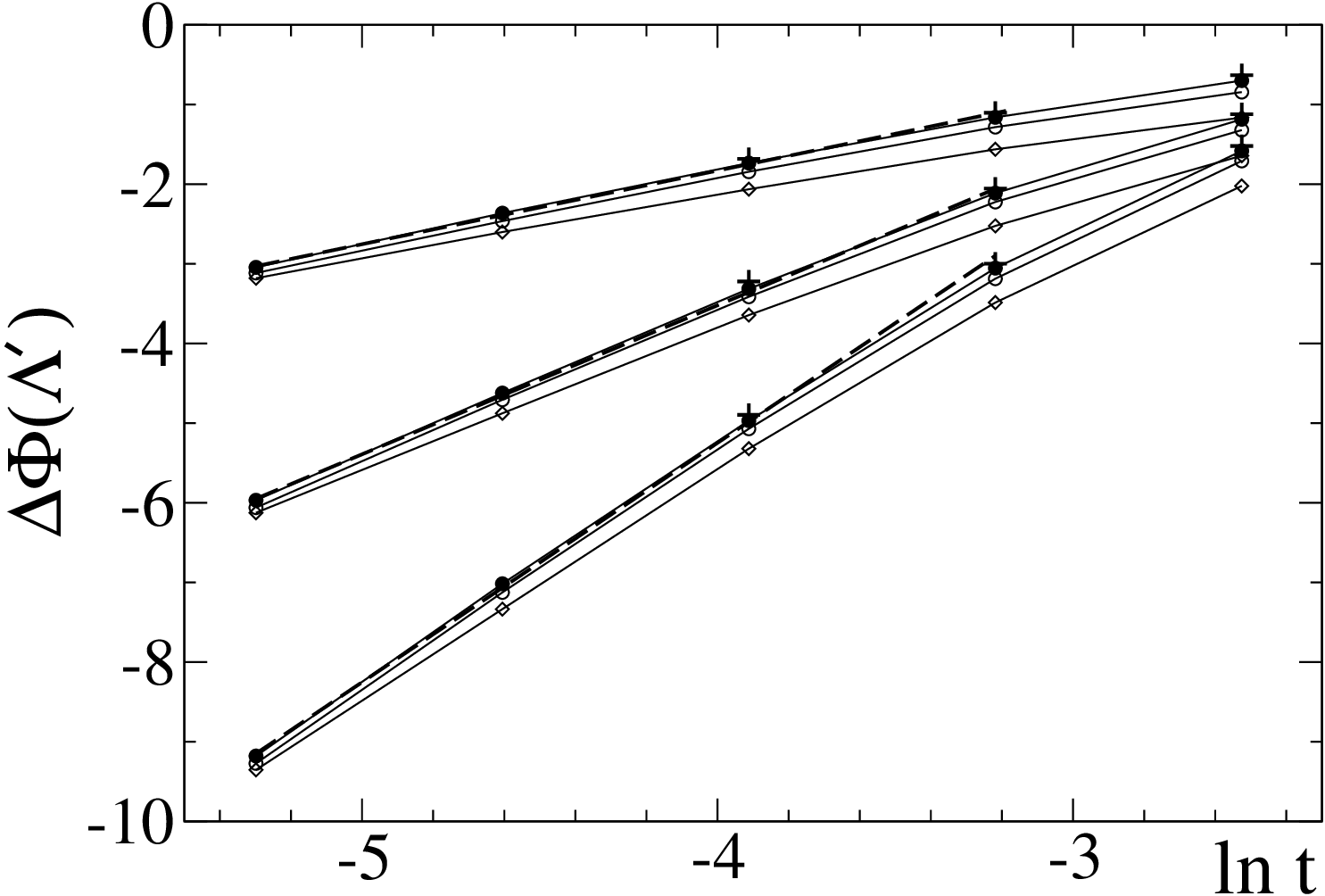}
\end{center}
\caption{The $\Delta \Phi(\Lambda')$ vs $\ln t$ plots at $tL=1.28$ (diamonds), $tL=2.56$ (empty circles),
$tL=5.12$ (solid circles) and $tL=10.24$ (pluses) for $\Lambda'=7 \pi/8$ (upper plots),
$\Lambda'=3 \pi/4$ (middle plots) and $\Lambda'=5 \pi/8$ (lower plots). The dashed straight lines 
are depicted to show that the plots at $tL=5.12$ are almost linear for small $t$ values.
Statistical errors are smaller than symbol size.}
\label{fig:Dphi}
\end{figure}

We have performed MC simulations for $\lambda=0.1$ at certain values of the reduced temperature,
$t = 1 - \beta/\beta_c = 0.08, 0.04, 0.02, 0.01, 0.005$, assuming $\beta_c=0.606479$
in accordance with the estimation in Sec.~\ref{sec:crp}. The error in this $\beta_c$ value
is as small as few times $10^{-7}$ and therefore is negligible in our analysis.
The simulations have been performed for lattice sizes $L=16, 32, 64, 128, 256, 512, 1024$ and $2048$
in order to evaluate the quantities $\Phi$ and $\Delta \Phi(\Lambda')$ for $tL= 1.28, 2.56, 5.12$ and $10.24$.
It corresponds to $L/\xi_{\mathrm{2nd}} \approx 6.9, 13.8, 27.6$ and $55.2$ at the largest $L$ values, where
$\xi_{\mathrm{2nd}}$ is the second moment correlation length, defined as in~\cite{HasRev},
i.~e., $\xi_{\mathrm{2nd}} = \sqrt{[(\chi/G(2\pi/L))-1]/ [4 \sin^2(\pi/L)]}$.
According to this, it can be expected that the results for $tL=5.12$ and $tL=10.24$ provide good
approximations for the thermodynamic limit, since  $L/\xi_{\mathrm{2nd}} \gg 1$ holds.
It is confirmed by the $\Phi$ vs $\ln t$ plots in Fig.~\ref{fig:Phi} and $\Delta \Phi(\Lambda')$ vs $\ln t$
plots in Fig.~\ref{fig:Dphi}, showing a fast convergence to the thermodynamic limit with increasing
of $tL$ at a fixed $t$.

These plots tend to become linear at large $tL$ and small $t$ values.
It indicates that $\Phi$ and $\Delta \Phi(\Lambda')$ have logarithmic singularities
in the thermodynamic limit at $t \to 0$. Moreover, it holds for $\Delta \Phi(\Lambda')$
at arbitrary $\Lambda'$, implying that $\partial G(k)/\partial \beta$ has the logarithmic
singularity at $t \to 0$ for any fixed non-zero $k$ in the thermodynamic limit.
Although we have tested only the $\langle 10 \rangle$ direction, this, obviously,
is true also for $\partial G({\bf k})/\partial \beta$ and 
$\partial G({\bf k})/\partial t = -\beta_c \partial G({\bf k})/\partial \beta$
at any fixed non-zero wave vector ${\bf k}$, since $G({\bf k})$
is a continuous function of ${\bf k}$ for $\mid {\bf k} \mid >0$.
Moreover, critical singularities are universal and, therefore, 
$\partial G({\bf k})/\partial t$ exhibits such logarithmic singularity both in the lattice model 
and in the continuous model.
The numerical analysis alone cannot provide a real proof that the discussed
here singularities are exactly logarithmic. 
On the other hand, the singularity of $\partial G({\bf k})/\partial t$ and the related asymptotic singularities could not be 
only approximately logarithmic, since $\partial G({\bf k})/\partial t$
contributes to specific heat~(\ref{eq:Cvs2}) (or~(\ref{eq:CVsing}), where
$G({\bf k}) - G^*({\bf k}) = t \, \partial G({\bf k})/\partial t$ holds at $t \to 0$ and ${\bf k \ne 0}$),
but the singularity of specific heat is known to be exactly logarithmic for the models of 2D Ising universality class,
including the scalar 2D $\varphi^4$ model.

According to the scaling hypothesis~(\ref{eq:sc1}), the asymptotic small-$t$ behavior of $\partial G({\bf k})/\partial t$
can be reached only at $\xi \sim 1/k$, i.~e., at $t \sim k$ in our case.
Therefore, the asymptotic  small-$t$ behavior of $\Delta \Phi(\Lambda')$ in the thermodynamic limit is reached with a given accuracy at $t<t^*$, 
where $t^* \to 0$ at $\Lambda' \to 0$. It is consistent with the fact that the linearity of the 
$\Delta \Phi(\Lambda')$ vs $\ln t$ plot is better for $\Lambda'=7\pi/8$ and $\Lambda'=3 \pi/4$ than for $\Lambda'=5 \pi/8$ in Fig.~\ref{fig:Dphi}.

The logarithmic singularity of $\partial G({\bf k})/\partial t$ implies that 
$\Delta \Phi(\Lambda')=\Phi(\pi)-\Phi(\Lambda')$  
behaves as $\sim \ln t$ in the thermodynamic limit at $t \to 0$, as a result of
an integration of $k \, \partial G({\bf k})/\partial t$ over $\Lambda' < \mid {\bf k} \mid < \Lambda$
(according to~(\ref{eq:DPhi}) at $L \to \infty$).
The same is true for $\Delta C_V^{sing}(\Lambda')=C_V^{sing}-C_V^{sing}(\Lambda')$ with the 
singular part of specific heat $C_V^{sing}$
given by~(\ref{eq:Cvs2}), $C_V^{sing}(\Lambda')$ being the contribution of the integration region $k < \Lambda'$.
Hence we find $C_V^{sing}(\Lambda') \sim \ln t$, using the fact that
$C_V^{sing} \sim \ln t$ holds 
in the actual 2D $\varphi^4$ model.
Since $C_V^{sing}(\Lambda')$ in the theorem is defined as the leading singular contribution of the $k<\Lambda'$ region at $t \to 0$, represented in powers
of $\xi$ and $\ln \xi$, we have $C_V^{sing}(\Lambda') \propto \ln \xi$. Consequently, the condition of the theorem 
$C_V^{sing}(\Lambda') = A(\Lambda') \, (\ln \xi)^{\lambda} \xi^{\alpha/\nu}$ is satisfied
here with $\lambda=1$ and $\alpha=0$. 

\begin{figure}
\begin{center}
\includegraphics[width=0.5\textwidth]{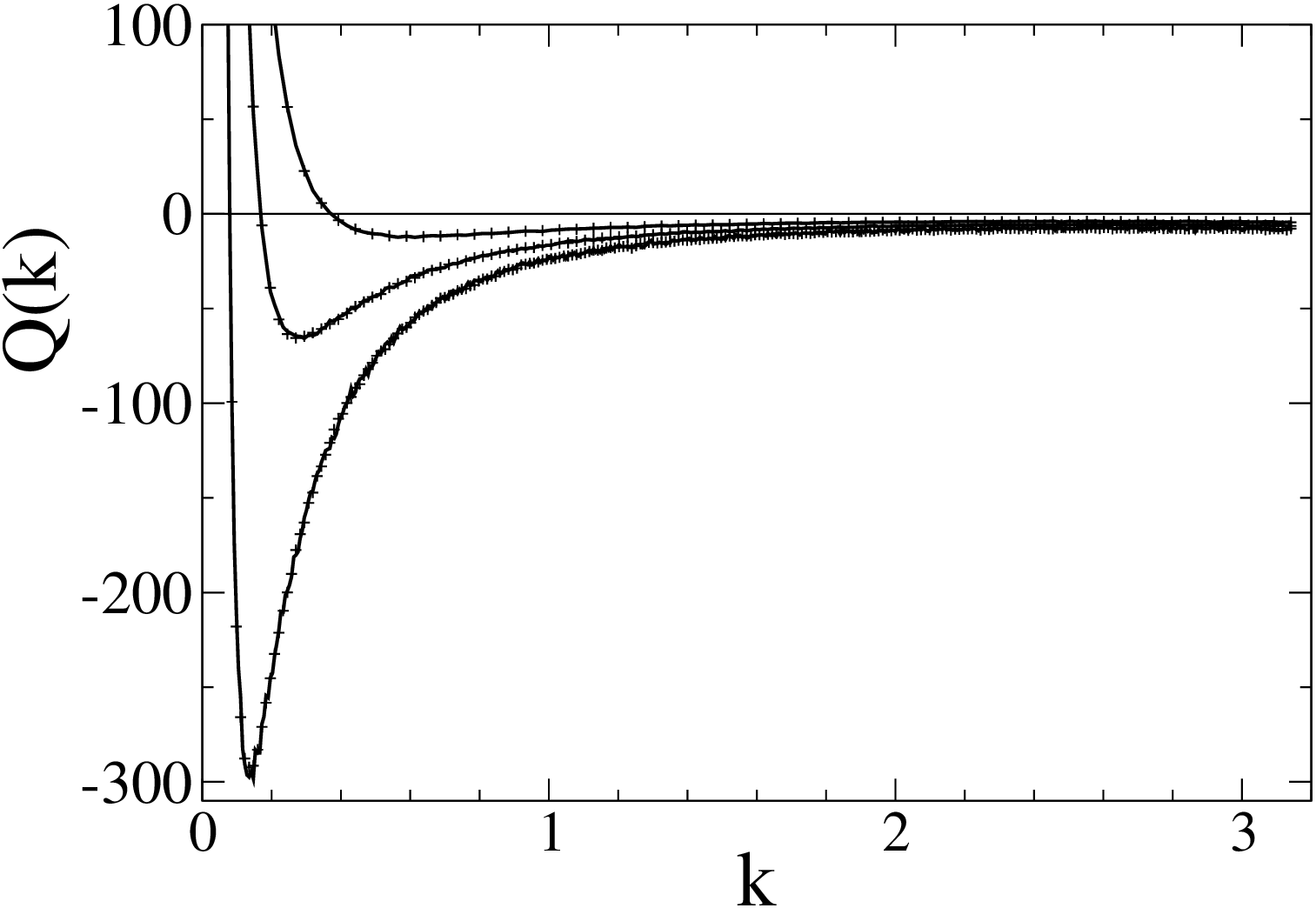}
\end{center}
\caption{$Q(k)=k \, \partial G(k)/\partial \beta$ vs $k$ plots at $t=0.02$ (upper curve),
$t=0.01$ (middle curve) and $t=0.005$ (lower curve). The results for $tL=5.12$ are shown by curves,
whereas those for $tL=2.56$ -- by pluses. The pluses lie practically on the top of curves,
showing that the thermodynamic limit is reached  with a high accuracy at $tL =5.12$.}
\label{fig:Gatv}
\end{figure}

As discussed before,
the long--wavelength (small-$k$) contributions, i.~e., $\Phi(\Lambda')$ and $C_V^{sing}(\Lambda')$ at small $\Lambda'$ values,
have similar singularities. The logarithmic singularity of $C_V^{sing}(\Lambda')$ thus means 
that $\Phi(\Lambda') = \mathcal{B}_1(\Lambda') \ln t$ holds with some coefficient $\mathcal{B}_1(\Lambda')$  for small 
cut-off parameter $\Lambda'$  in the thermodynamic limit
at $t \to 0$. Moreover, since $\partial G(k)/\partial t$ has a logarithmic singularity at any fixed positive $k$ 
in this limit, the asymptotic relation
$\Phi(\Lambda') = \mathcal{B}_1(\Lambda') \ln t$ can be extended (by integrating $k \partial G(k)/\partial \beta$ over $k$)
to any finite value of $\Lambda'$ not exceeding $\Lambda$.
We have also a similar asymptotic relation $\Delta \Phi(\Lambda') = \mathcal{B}_2(\Lambda') \ln t$
for $\Delta \Phi(\Lambda')$, as consistent with our earlier statements. It yields 
$\Phi = \mathcal{B} \ln t$ with $\mathcal{B}=\mathcal{B}_1(\Lambda')+\mathcal{B}_2(\Lambda')$
for $\Phi = \Phi(\Lambda') + \Delta \Phi(\Lambda')$ in the thermodynamic limit at $t \to 0$.
As an extra argument, the plots in Figs.~\ref{fig:Phi} and~\ref{fig:Dphi} provide a direct numerical
evidence that these relations and logarithmic singularities really hold true.
It is clear from Fig.~\ref{fig:Phi} that $\mathcal{B} <0$ holds, since the asymptotic slope of the plot (for $tL \to \infty$) is negative.
On the other hand, $\partial G(k)/\partial \beta$ is negative at $t \to 0$ for any fixed $\Lambda'$ in the thermodynamic limit,
according to the scaling behavior shown in Fig.~\ref{fig:Gatv}, where $\partial G(k)/\partial \beta <0$
holds for $k>k^*(t)$ with $k^*(t)$ tending to zero approximately as $\propto t$ at $t \to 0$. 
It means that $\mathcal{B}_2(\Lambda')>0$ holds.
In such a way, we have $\mathcal{B}_1(\Lambda') = \mathcal{B} - \mathcal{B}_2(\Lambda') < \mathcal{B}$ and
thus $\lim_{\Lambda' \to 0} \mathcal{B}_1(\Lambda') \ne 0$, i.~e., the long--wavelength contribution to $\Phi$
is relevant. Consequently, the corresponding (similar) long--wavelength contribution to $C_V^{sing}$ is also relevant,
implying that the condition of the theorem $\lim_{\Lambda' \to 0} A(\Lambda') \ne 0$ is satisfied.

\section{Monte Carlo analysis}
\label{sec:MCanal}

\subsection{Relations of finite--size scaling}
\label{sec:ancorr}

According to the finite--size scaling theory, susceptibility can be represented as
\begin{equation}
 \chi= L^{\gamma/\nu} \left( f_0(L/\xi) + f_1(L/\xi)L^{-\omega} + \cdots \right) + \chi_{\mathrm{anal}}(t,L) \;,
\label{eq:chis}
 \end{equation}
where $f_i(L/\xi)$ are scaling functions, $\omega$ is the leading correction--to--scaling exponent,
and $\chi_{\mathrm{anal}}(t,L)$ is the analytical background contribution. The singular part of 
$\partial U/\partial \beta$ can be represented in a similar way as that of $\chi$ with the 
scaling exponent $1/\nu$ instead of $\gamma/\nu$. The scaling argument $L/\xi$ tends to a constant
at $\beta = \widetilde{\beta}_c(L)$ and $L \to \infty$. Consequently,
if the actual $\varphi^4$ model is described by the same critical exponents
$\gamma = 7/4$ and $\nu = 1$ as the 2D Ising model, then $\chi/L^{7/4}$ and
$(\partial U/\partial \beta)/L$
at $\beta = \widetilde{\beta}_c(L)$ tend to some nonzero constants at $L \to \infty$. The data in Tabs.~\ref{tab1}
to~\ref{tab6} are consistent with this idea. The $L$--dependence of $\chi/L^{7/4}$ and
$(\partial U/\partial \beta)/L$ is caused by corrections to scaling, including those coming from
the analytic background term, if it exists. Thus, we have
\begin{eqnarray}
 \chi/L^{7/4} &=& a_0 + \sum\limits_{k \ge 1} a_k L^{-\omega_k} \label{eq:chi} \;, \\
 \frac{1}{L} \, \frac{\partial U}{\partial \beta}  &=& b_0 + \sum\limits_{k \ge 1} b_k L^{-\omega_k} \label{eq:uatv}
\end{eqnarray}
for large $L$ at $\beta = \widetilde{\beta}_c(L)$, where $a_k$ and $b_k$ are expansion coefficients and $\omega_k$ are correction--to--scaling
exponents. The existence of trivial corrections to scaling with integer $\omega_k$ is expected, since such 
corrections appear in the 2D Ising model.

The numerical analysis in~\cite{CGNP11} indicates that the susceptibility of the 2D Ising model
on various lattices contains logarithmic corrections, coming from the ``short--distance'' contribution of the form
\begin{equation}
 B^{\mathrm{lattice}} = \sum\limits_{q=0}^{\infty} \sum\limits_{p=0}^{[\sqrt{q}]} b^{(p,q)} (\ln \mid t \mid)^p t^q \;.
\label{eq:log}
 \end{equation}
These terms with $p>0$ in~(\ref{eq:log}) represent a correction of order $O(t \ln \mid t \mid)$,
which is a quantity of order $O(\ln L /L)$ in the finite--size scaling regime $t \sim 1/L$. These are
high--order correction terms, which are not included in our fits, since the leading of them is by a 
factor $\sim L^{-11/4} \ln L$ smaller than the susceptibility $\chi$ at $L \to \infty$.

The singular terms in~(\ref{eq:chis}), which are not related to~(\ref{eq:log}), will
be further referred as the ``long--distance'' singular contributions.
According to~\cite{CGNP11}, these are representable by integer correction exponents
in~(\ref{eq:chis}) at $L/\xi \to \infty$ in the case of the 2D Ising model.
This is usually expected to be true also at finite values of $L/\xi$.
Thus, if corrections to scaling in the scalar 2D $\varphi^4$ model have such structure, then~(\ref{eq:chi})
contains corrections $a_1L^{-1}$, $a_2L^{-7/4}$, $a_3L^{-2}$ and corrections of higher orders.
Moreover, in this case we have $a_2=\chi_{\mathrm{anal}}(0,\infty)$, so that the coefficient $a_2$ is independent of the 
particular choice of $\widetilde{\beta}_c(L)$, i.~e., choice of $U$.
Following the analogy with 2D Ising model, one can expect
that $a_1$ vanishes at $\beta=\beta_c$ and, therefore, probably also at $U=U^*$.
In distinction from susceptibility, $\partial U / \partial \beta$ does not contain such a constant contribution, 
which comes from an analytical background term,
since $U$ is constant ($U=3$ at $\beta<\beta_c$ and $U=1$ at $\beta>\beta_c$)
at $\beta \ne \beta_c$ and $L \to \infty$. A constant contribution  
can, nevertheless, exist as a correction to the leading singular term.

\subsection{Preliminary finite--size scaling analysis of the data}
\label{sec:preliminary}

\begin{figure}
\begin{center}
\includegraphics[width=0.32\textwidth]{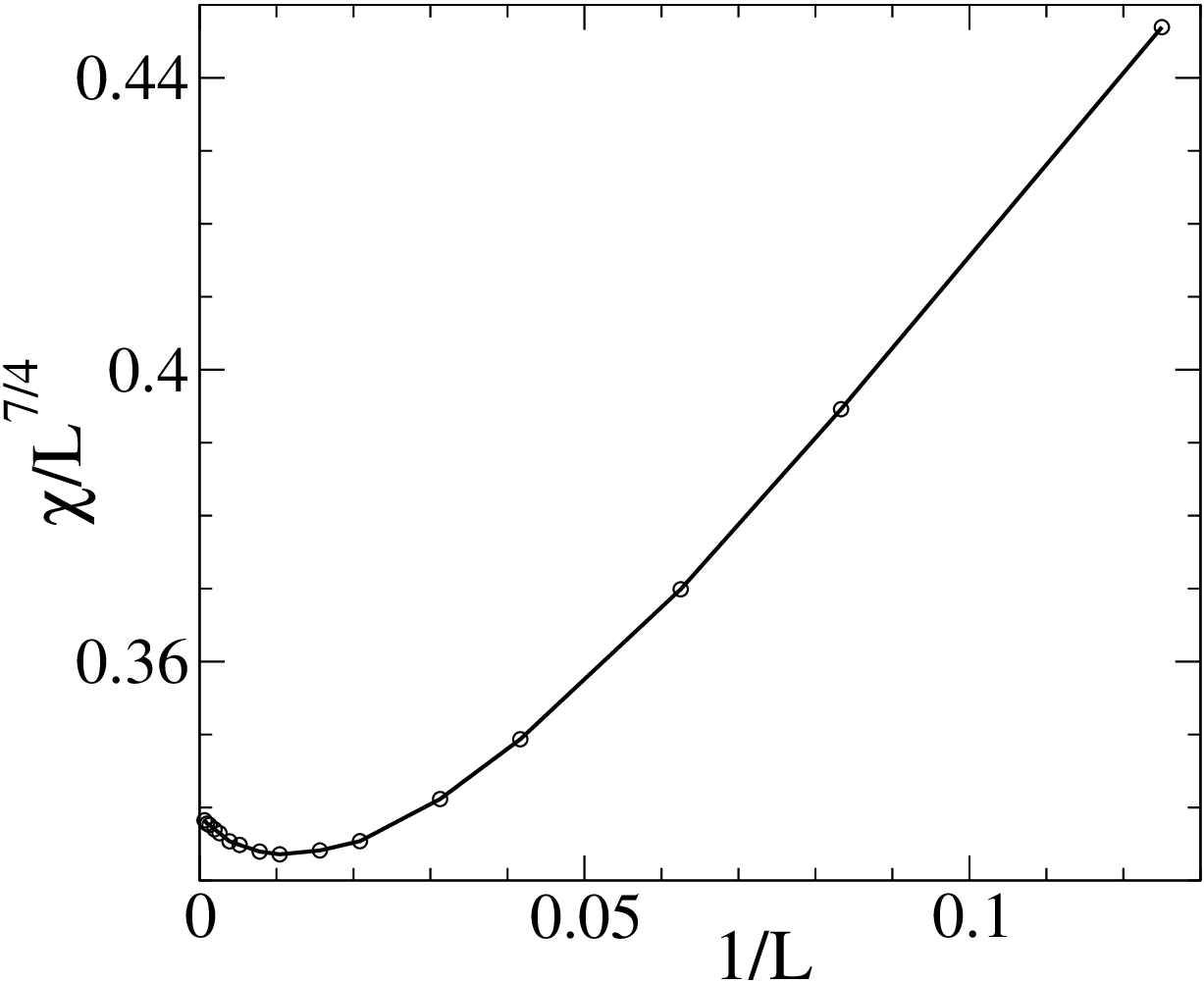}
\hfill
\includegraphics[width=0.32\textwidth]{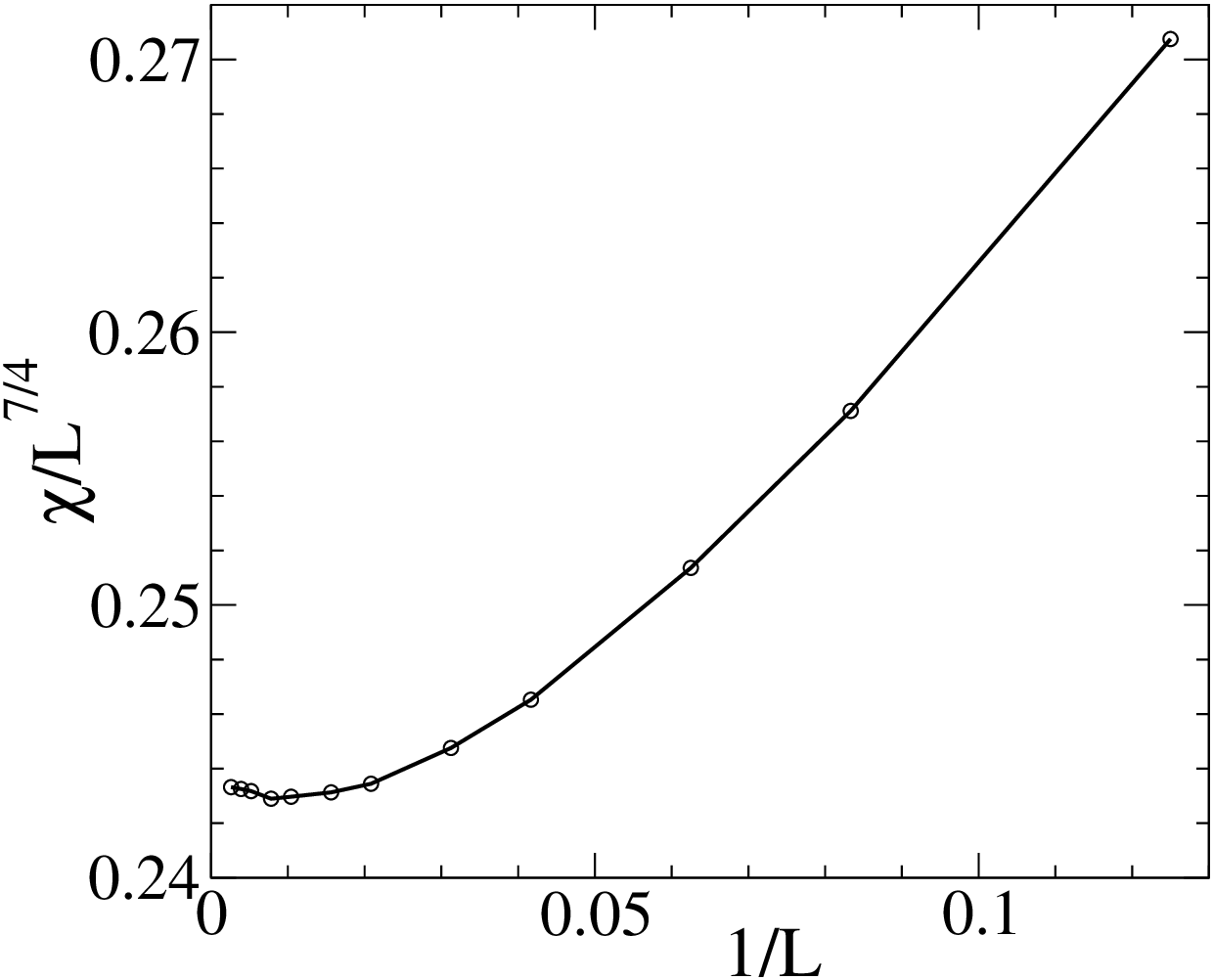}
\hfill
\includegraphics[width=0.32\textwidth]{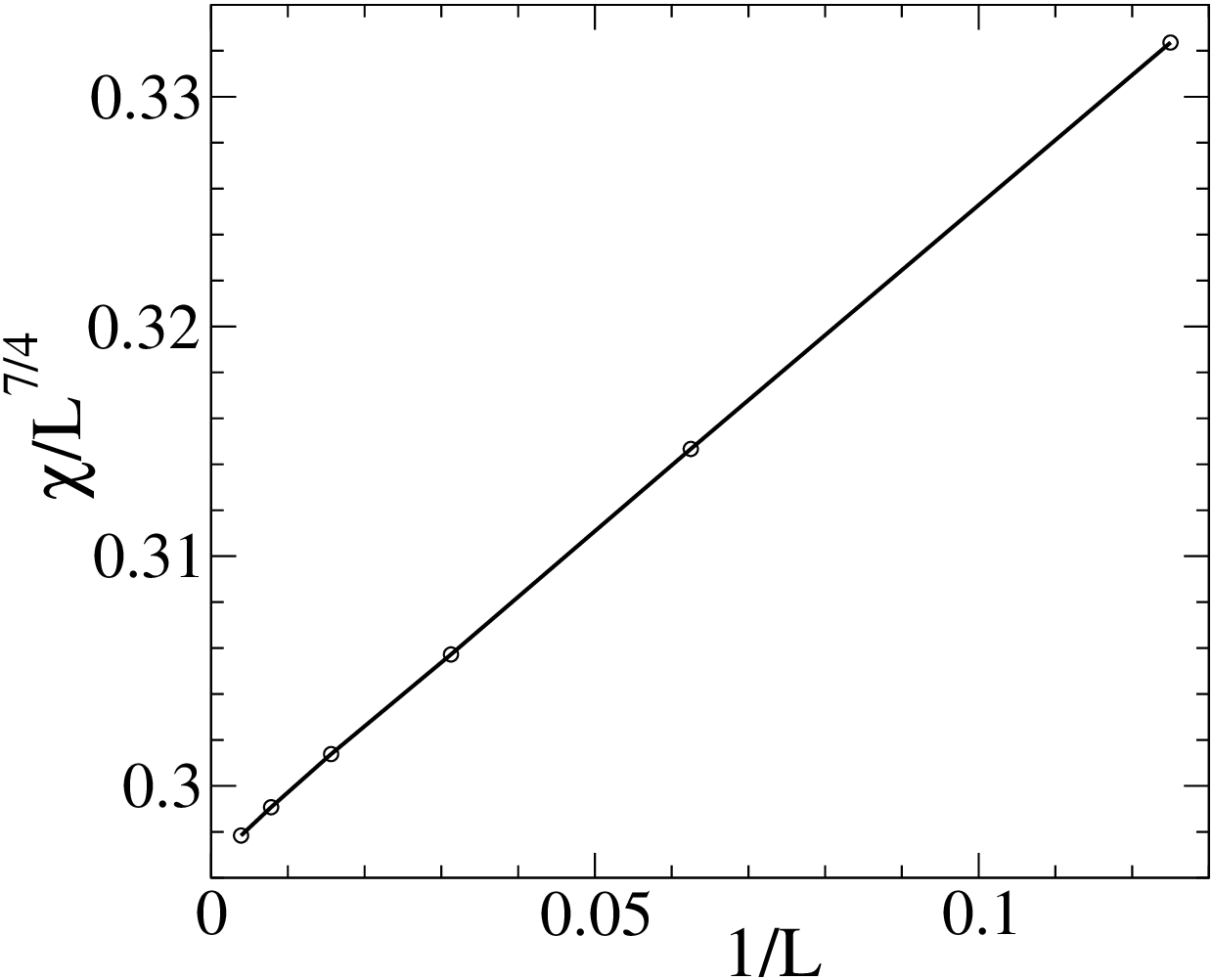} \\
\vspace*{3mm}
\includegraphics[width=0.32\textwidth]{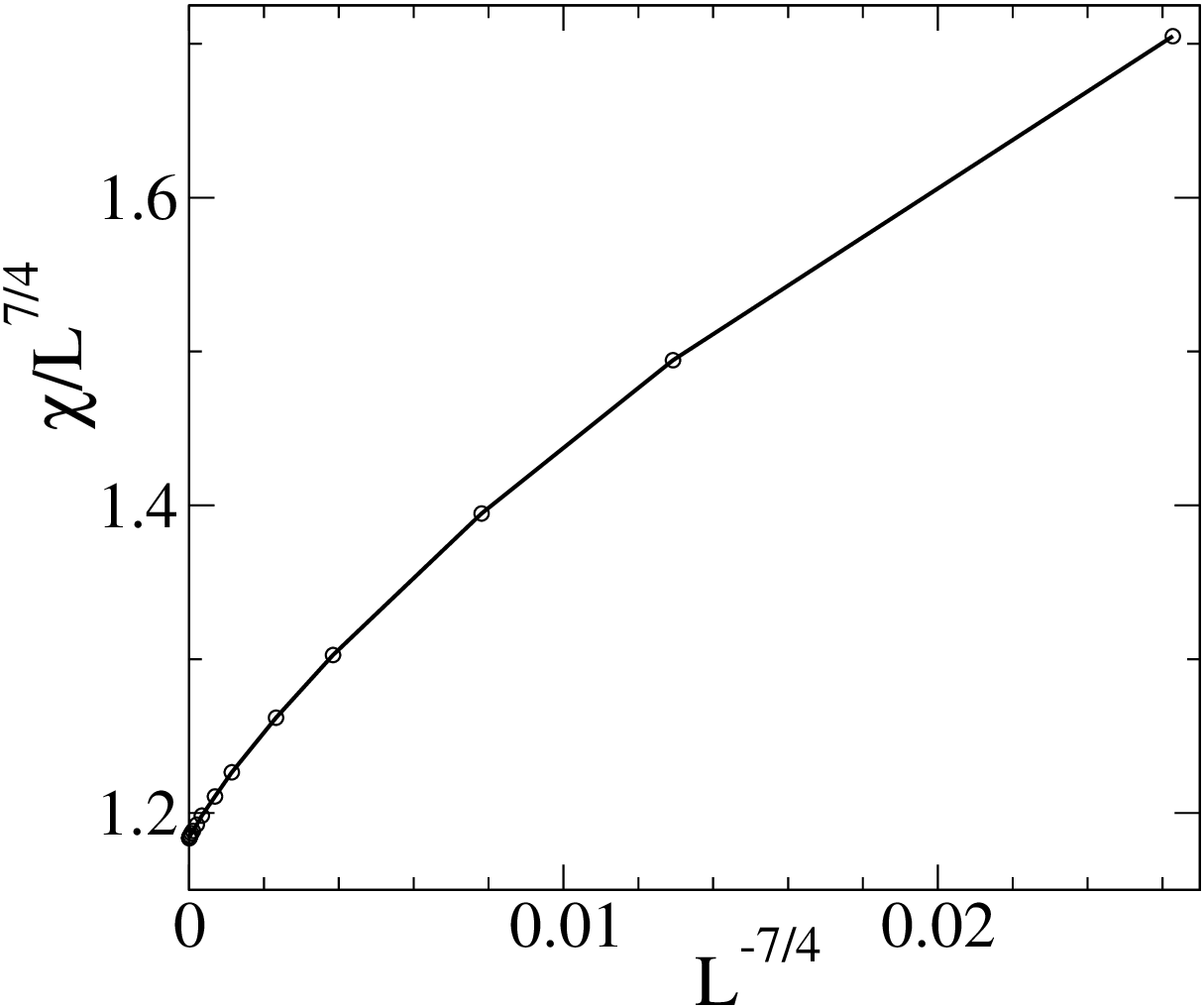}
\hfill
\includegraphics[width=0.32\textwidth]{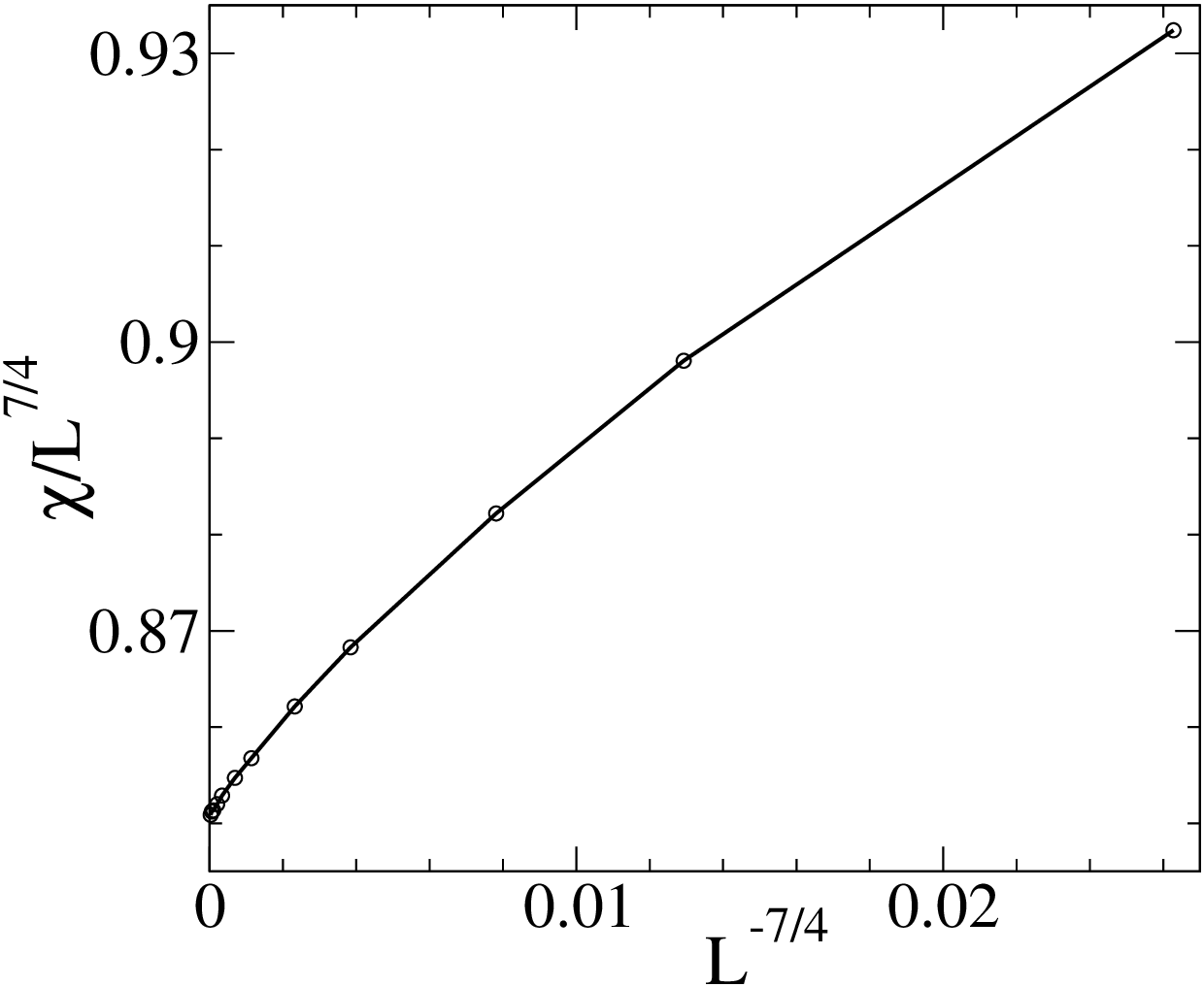}
\hfill
\includegraphics[width=0.32\textwidth]{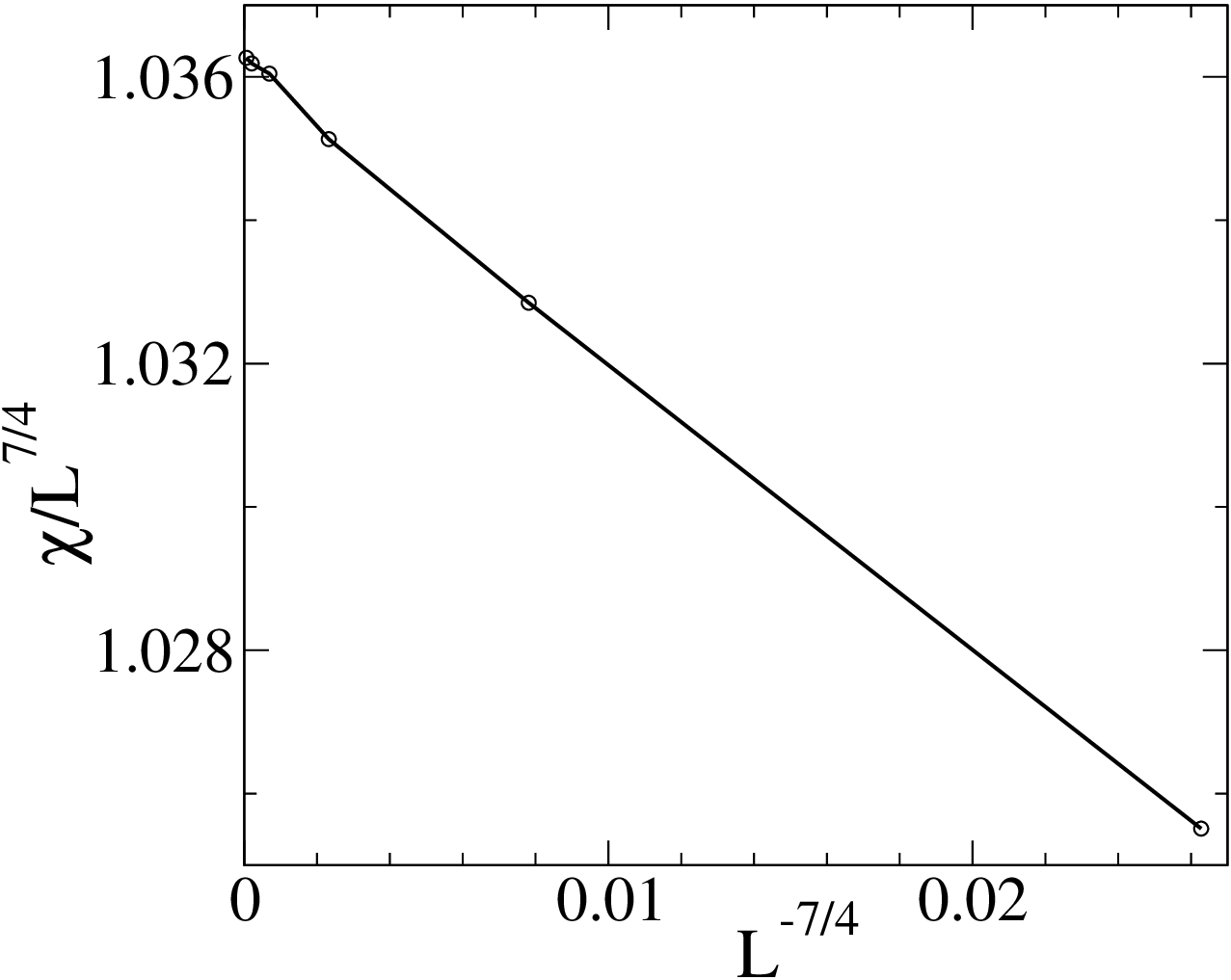}
\end{center}
\caption{The $\chi/L^{7/4}$ vs $1/L$ plots for $U=2$ (top) and 
$\chi/L^{7/4}$ vs $L^{-7/4}$ plots for $U=1.1679229 \approx U^*$ (bottom)
at $\lambda=0.1$ (left), $\lambda = 1$ (middle) and $\lambda = 10$ (right).
The range of sizes $L \ge 8$ is shown. 
Statistical errors are smaller than the symbol size.}
\label{chi}
\end{figure}

Summarizing the discussion in Sec.~\ref{sec:ancorr}, we conclude that
the $\chi/L^{7/4}$ vs $1/L$ plots are asymptotically linear at $L \to \infty$ for $U \ne U^*$
in the case of the Ising scenario, i.~e.,
if the structure of corrections to scaling in the scalar 
2D $\varphi^4$ model is similar to that one expected in the 2D Ising model.
The asymptotic linearity of the  
$\chi/L^{7/4}$ vs $L^{-7/4}$ plots can be expected in the special case of $U=U^*$.
Our MC results support the idea that this really corresponds to the Ising
scenario, since the $\chi/L^{7/4}$ vs $L^{-7/4}$ rather than $\chi/L^{7/4}$ vs $L^{-1}$
plot is almost linear for $U=1.1679229 \approx U^*$ at $\lambda=10$, i.~e., close
to the Ising limit. The discussed here plots are shown in Fig.~\ref{chi}, using the $1/L$ scale
for $U=2$ and $L^{-7/4}$ scale for $U \approx U^*$.
The plots  at $\lambda=0.1$ and $\lambda=1$ are remarkably nonlinear, whereas those at $\lambda=10$ look more linear.
The latter, however, is not surprising, since the Ising limit is approached at large values of $\lambda$. 
The best linearity is observed at $\lambda=10$ and $U=2$, where
the $\chi^2/\mathrm{d.o.f.}$ of the linear fit within $8 \le L \le 256$ is $2.33$. 

\begin{figure}
\begin{center}
\includegraphics[width=0.32\textwidth]{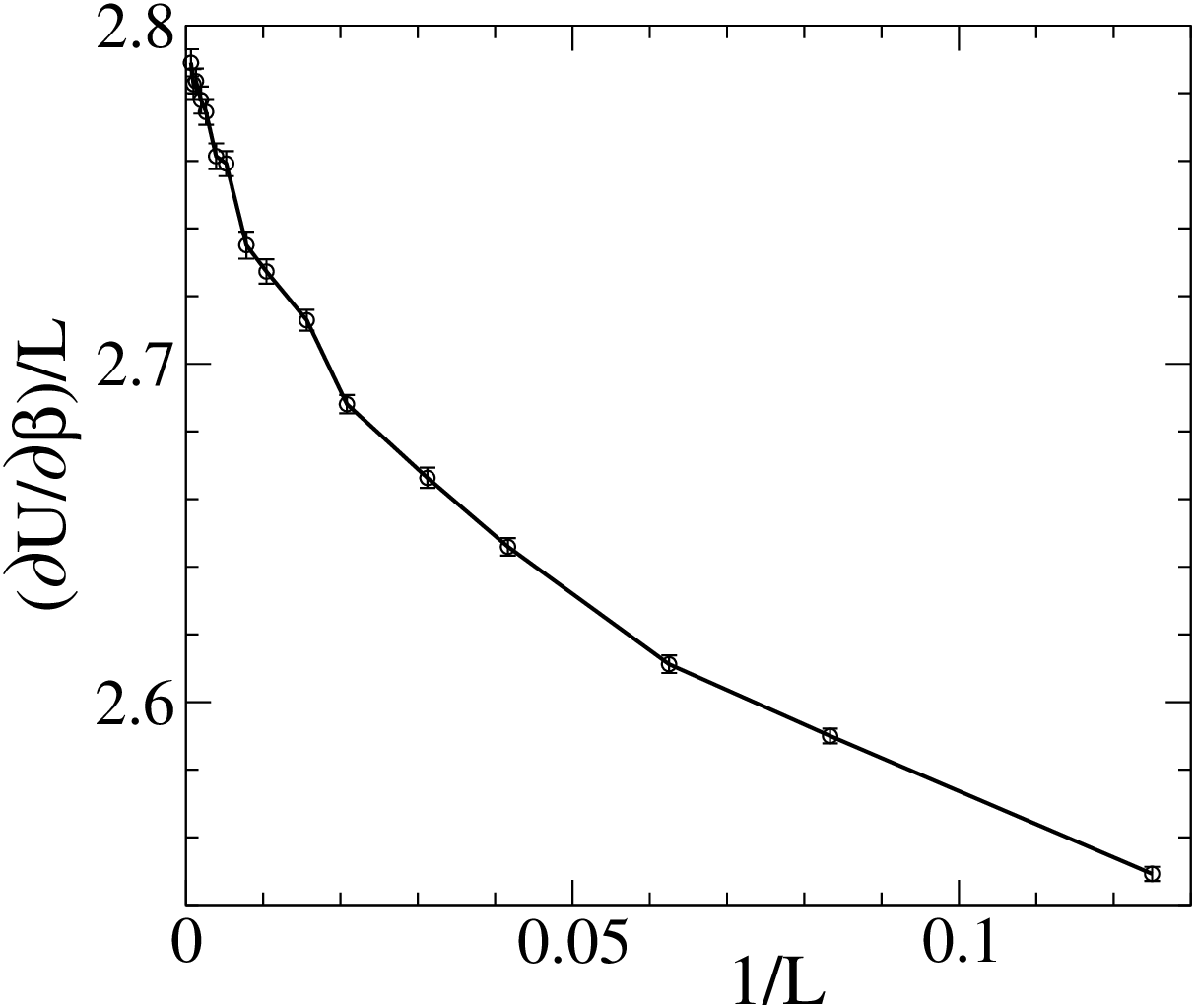}
\hfill
\includegraphics[width=0.32\textwidth]{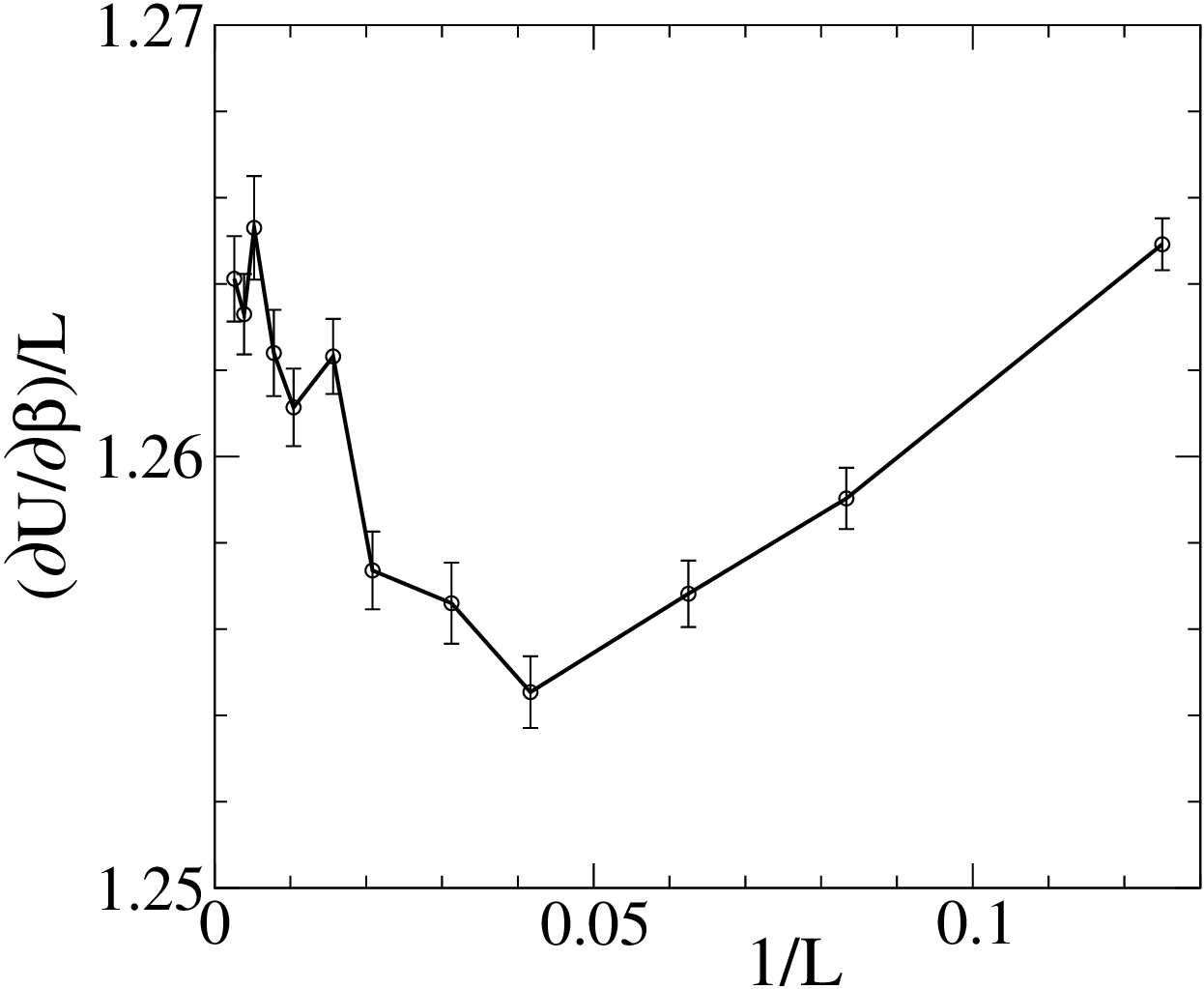}
\hfill
\includegraphics[width=0.32\textwidth]{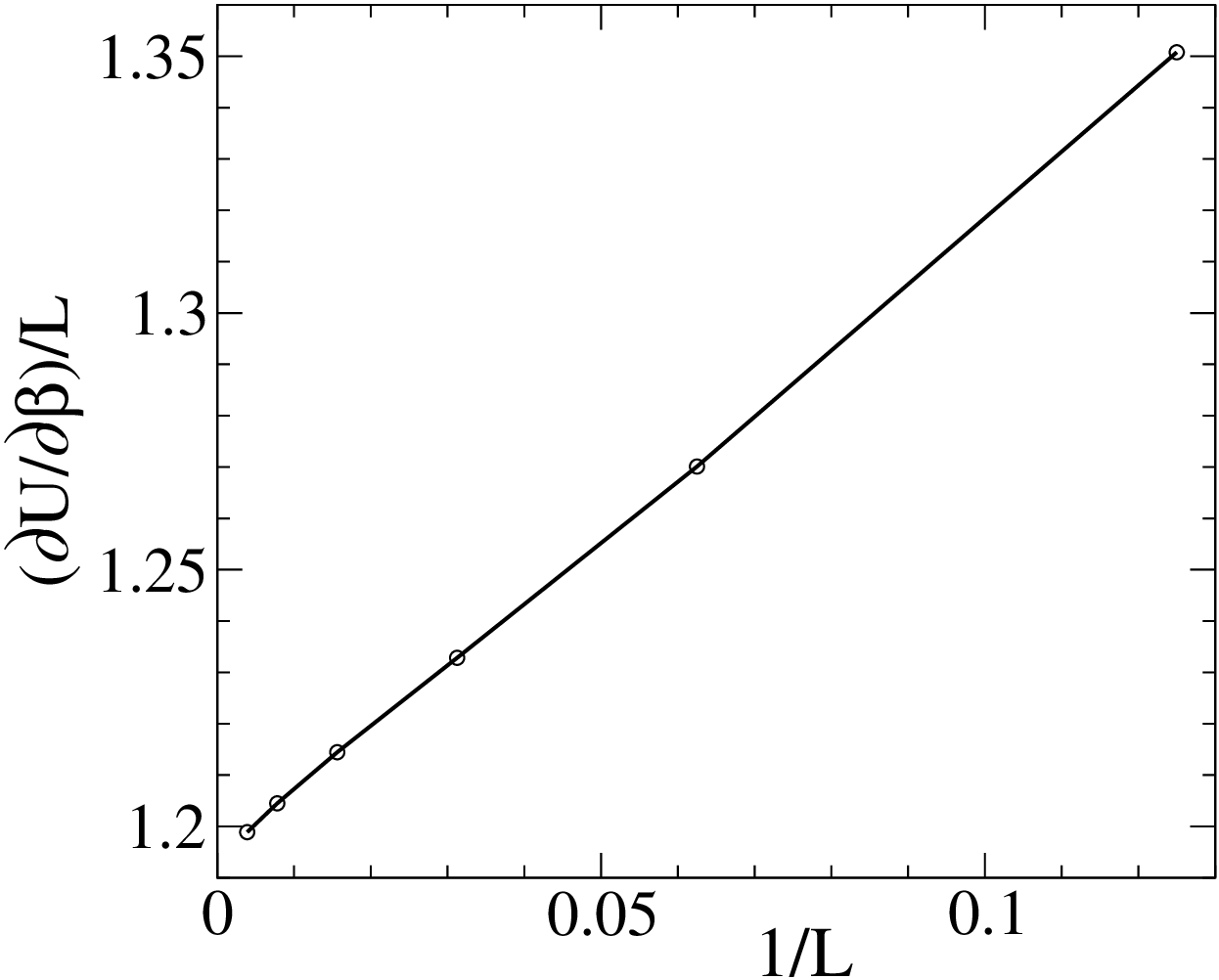} \\
\end{center}
\caption{The $-(\partial U/\partial \beta)/L$ vs $1/L$ plots for $U=2$ 
at $\lambda=0.1$ (left), $\lambda = 1$ (middle) and $\lambda = 10$ (right).
The range of sizes $L \ge 8$ is shown. 
Statistical errors at $\lambda=10$ are smaller than the symbol size.}
\label{U2}
\end{figure}

\begin{figure}
\begin{center}
\includegraphics[width=0.32\textwidth]{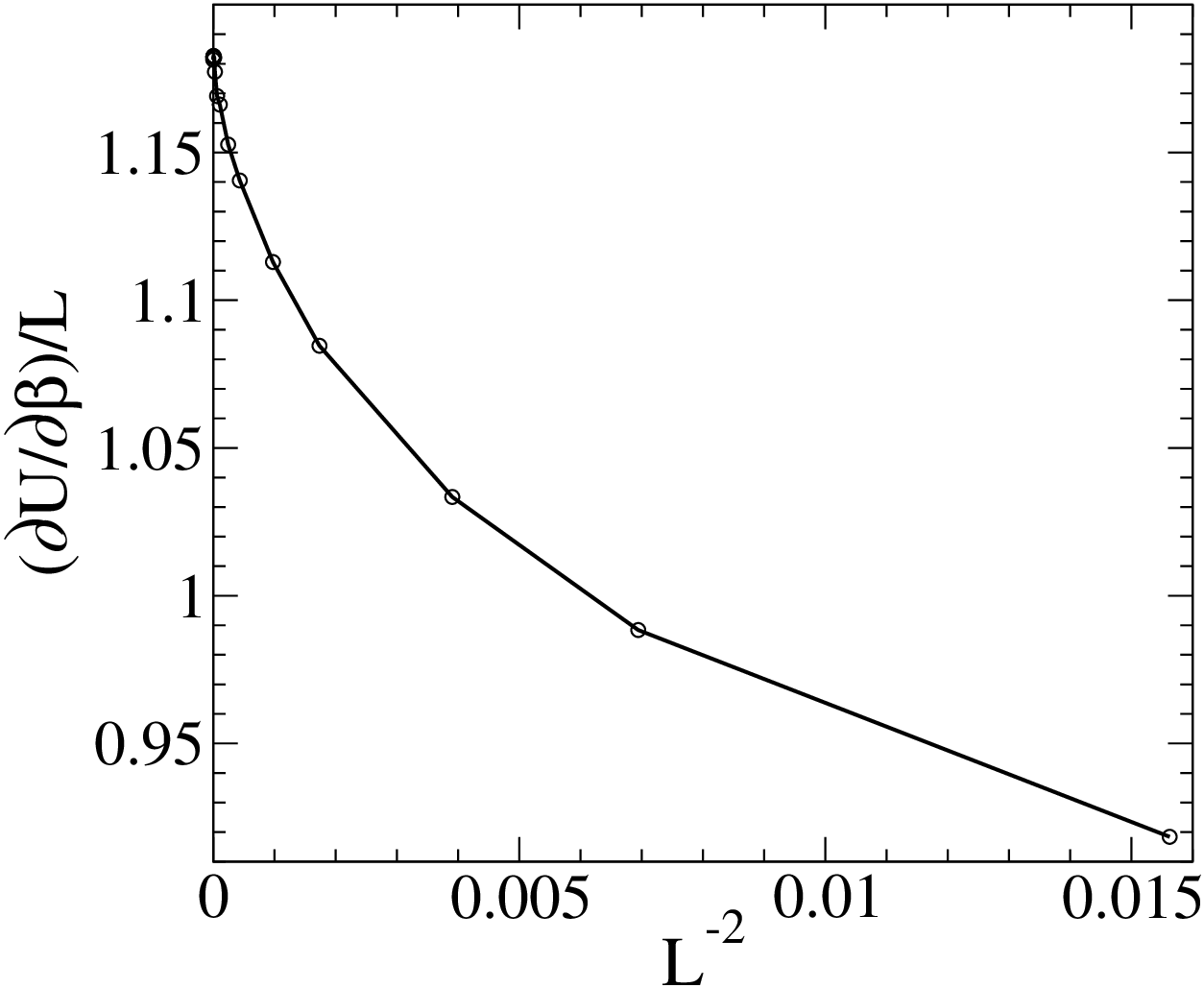}
\hfill
\includegraphics[width=0.32\textwidth]{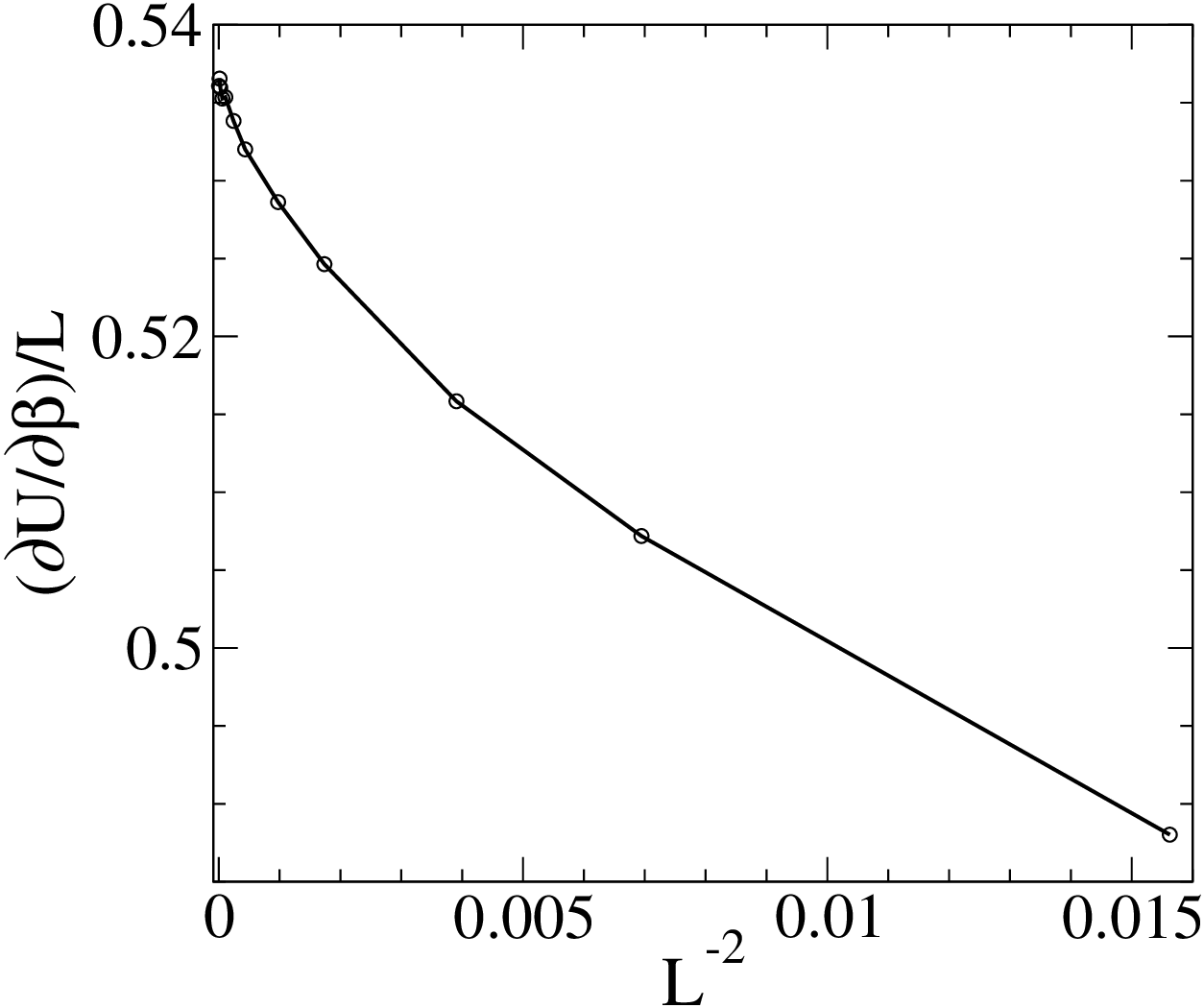}
\hfill
\includegraphics[width=0.32\textwidth]{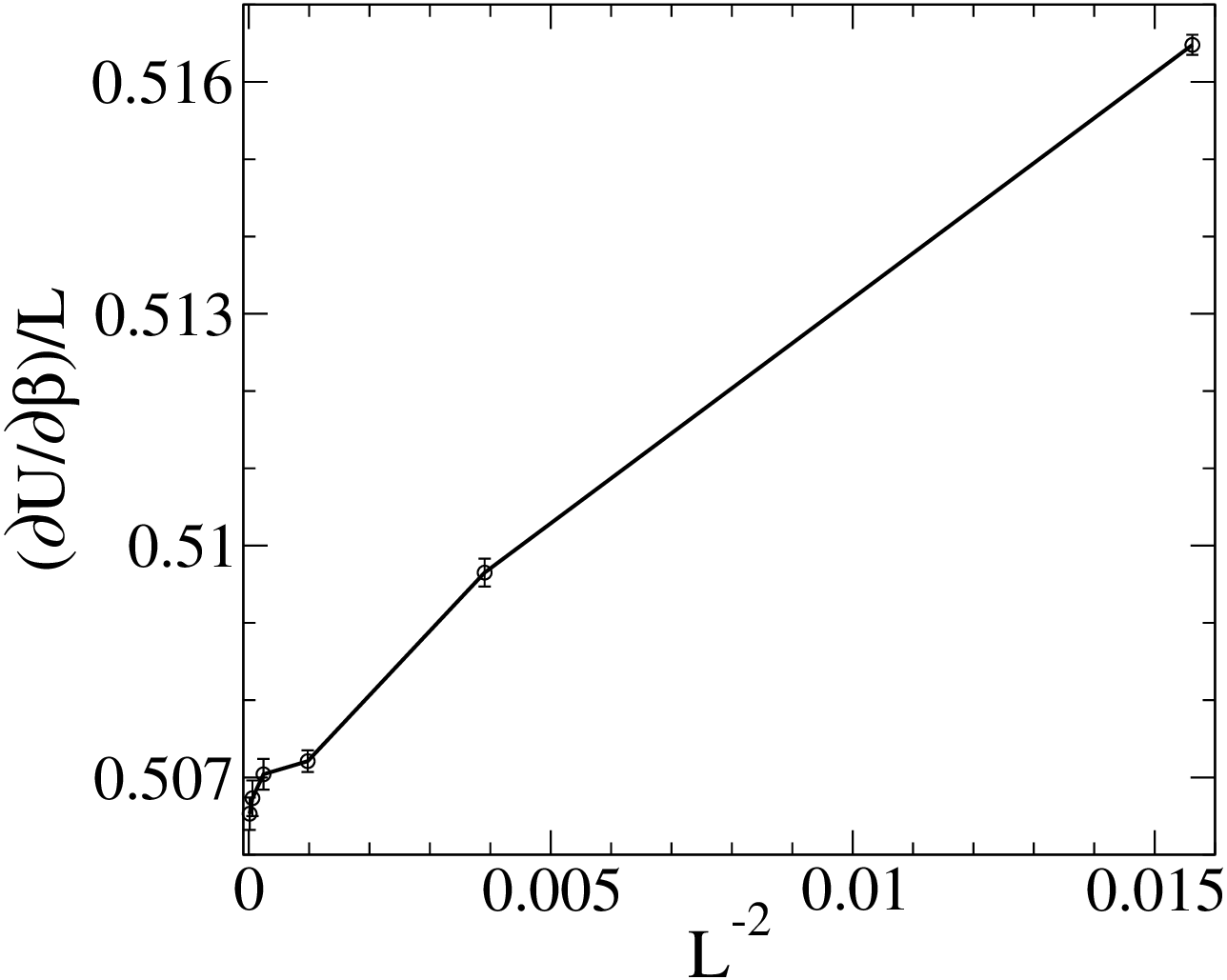} \\
\vspace*{3mm}
\includegraphics[width=0.32\textwidth]{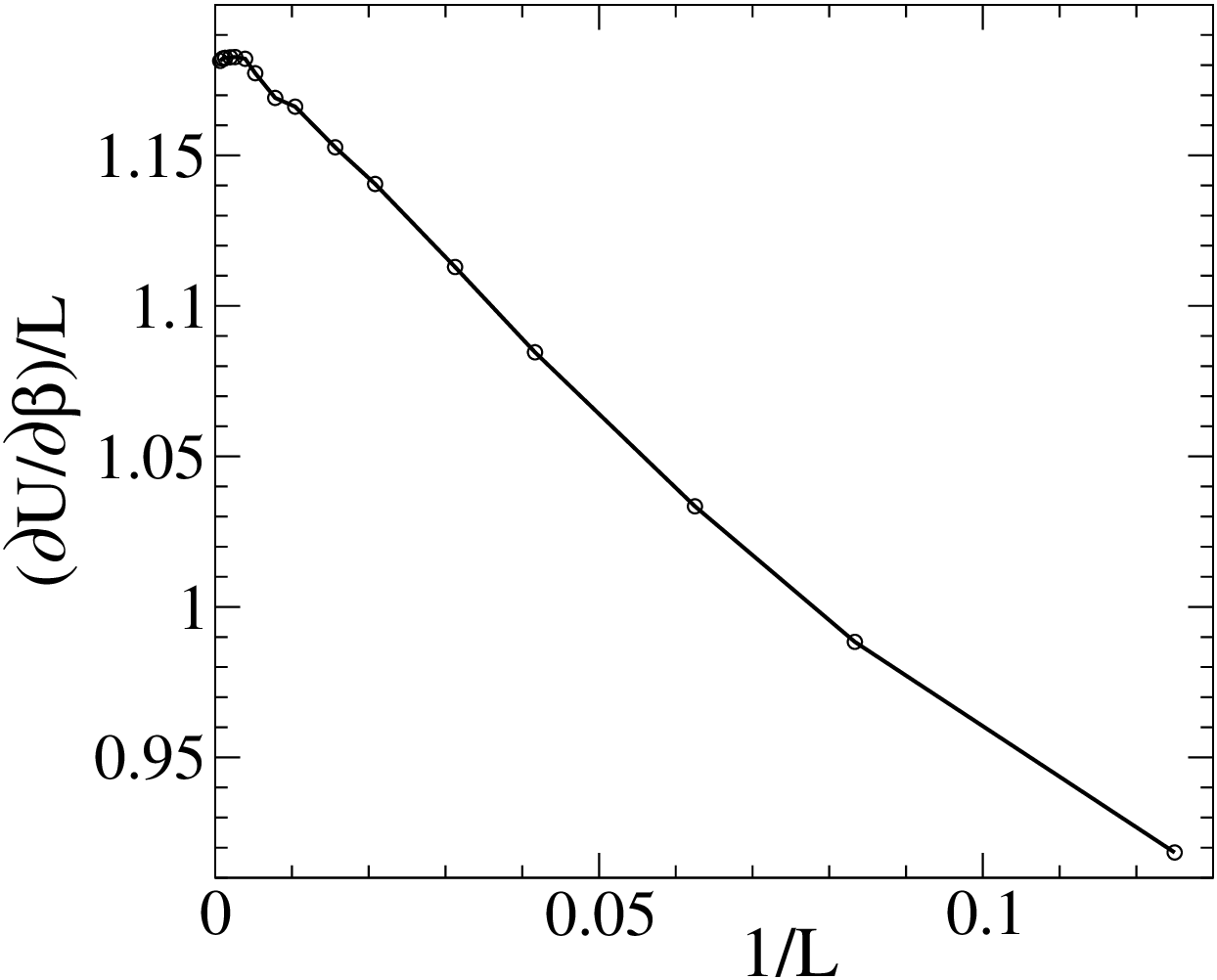}
\hfill
\includegraphics[width=0.32\textwidth]{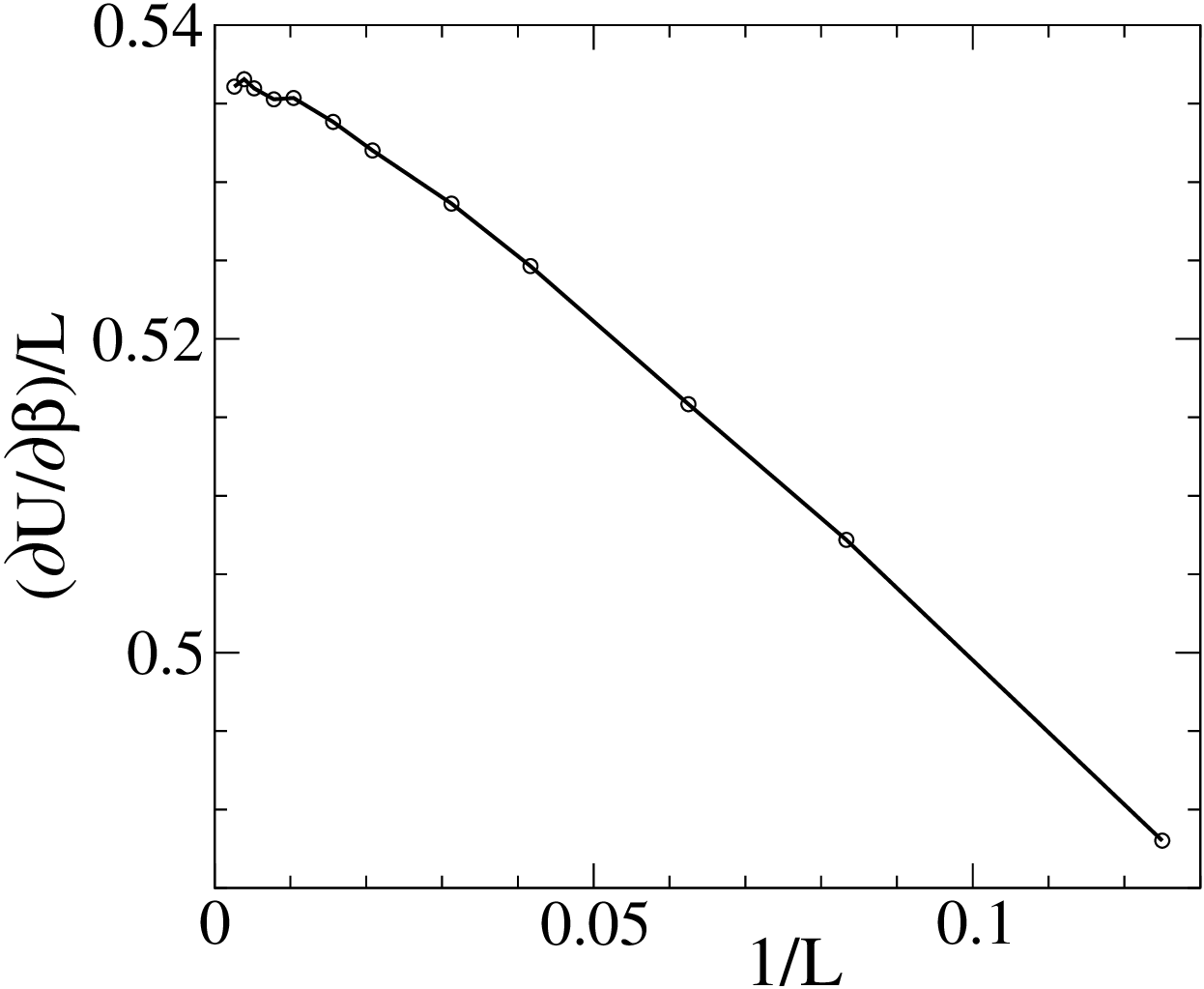}
\hfill
\includegraphics[width=0.32\textwidth]{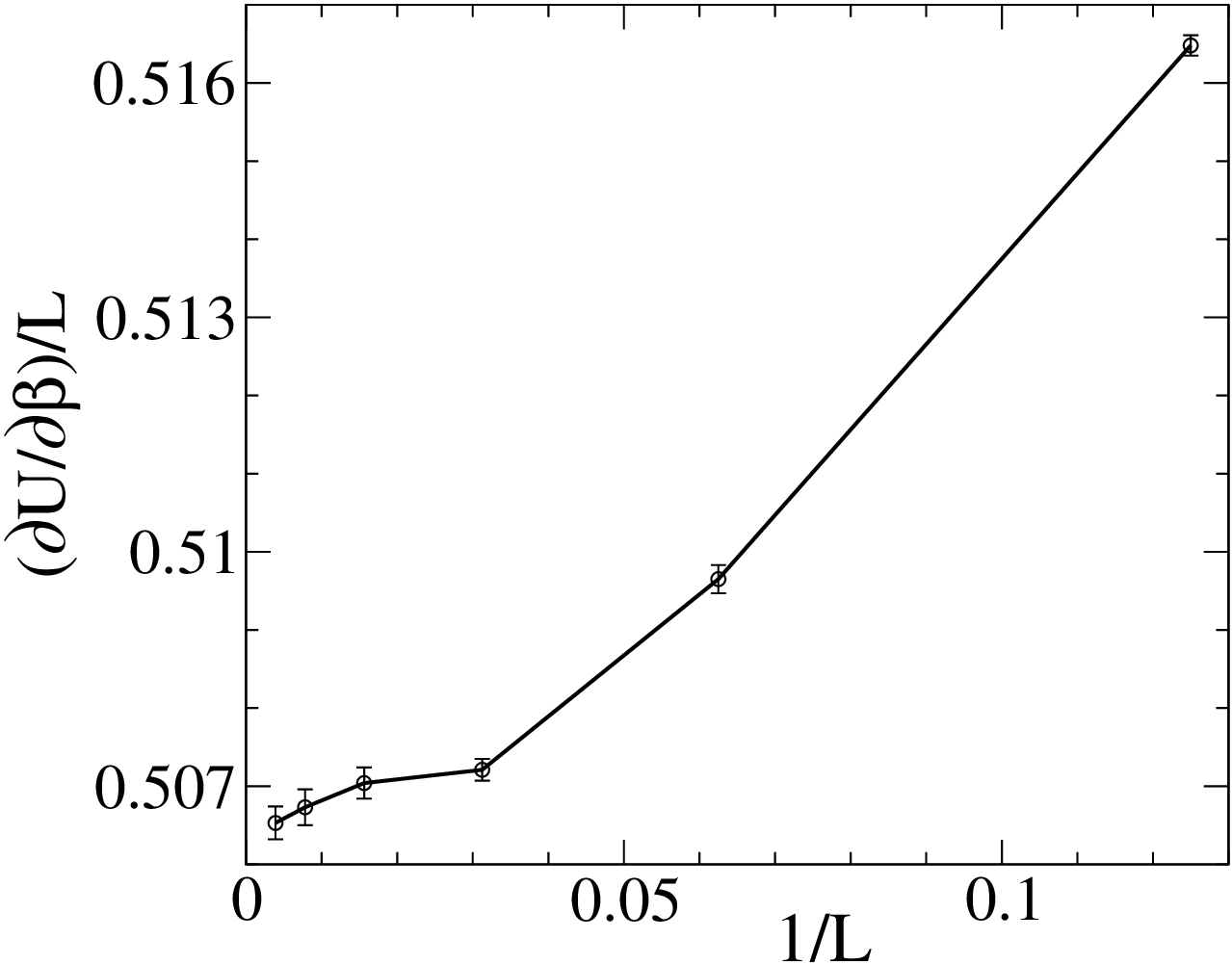}
\end{center}
\caption{The $-(\partial U/\partial \beta)/L$ plots for $U=1.1679229 \approx U^*$
depending on $1/L^2$ (top) and $1/L$ (bottom) 
at $\lambda=0.1$ (left), $\lambda = 1$ (middle) and $\lambda = 10$ (right).
The range of sizes $L \ge 8$ is shown. 
Statistical errors at $\lambda=0.1$ and $\lambda = 1$ are about the symbol size or smaller.}
\label{Uc}
\end{figure}

The asymptotic (large-$L$) linearity of the $-(\partial U/\partial \beta)/L$ vs $1/L$ plots can be 
expected according to the Ising scenario.
However, if the coefficient at $1/L$ vanishes for $U=U^*$ (as it is, probably, true for $\chi/L^{7/4}$ in the 2D Ising model), 
then the asymptotic linearity of $-(\partial U/\partial \beta)/L$ vs $L^{-2}$ plots
is expected in this particular case. 
The $-(\partial U/\partial \beta)/L$ vs $1/L$ plots for $U=2$ are shown in Fig.~\ref{U2},
whereas  $-(\partial U/\partial \beta)/L$ plots depending on $1/L$ and $1/L^2$ 
for $U=1.1679229 \approx U^*$ are shown in Fig.~\ref{Uc}. The best linearity is, again,
observed at $\lambda=10$ and $U=2$. However, even in this case the quality of the  $-(\partial U/\partial \beta)/L$ vs $1/L$
fit is low: $\chi^2/\mathrm{d.o.f.}=4.24$.

The nonlinearity of the plots at $\lambda=0.1$ and $\lambda=1$ in Figs.~\ref{chi} -- \ref{Uc} indicate that
nontrivial corrections to scaling with different exponents than those expected in the 2D Ising model could exist.
In this case, the approximate linearity of some of the plots at $\lambda=10$ can be easily explained
by the fact that the amplitudes of nontrivial correction terms vanish in the Ising limit $\lambda \to \infty$.
The analysis of this section is preliminary, since it
is based only on the evaluation of linearity of some plots. Nevertheless, we can expect from this analysis 
that the data for small values of $\lambda$ (such as $\lambda=0.1$), where the nonlinearity of these plots is more pronounced,
give the best chance to identify nontrivial correction terms, if they really exist.
Due to this reason, we have extended simulations up to $L=1536$ at $\lambda=0.1$ and have performed a refined 
analysis of the data in this case -- see Sec.~\ref{sec:estex}.

\subsection{Estimation of correction exponents}
\label{sec:estex}

\begin{figure}
\begin{center}
\includegraphics[width=0.5\textwidth]{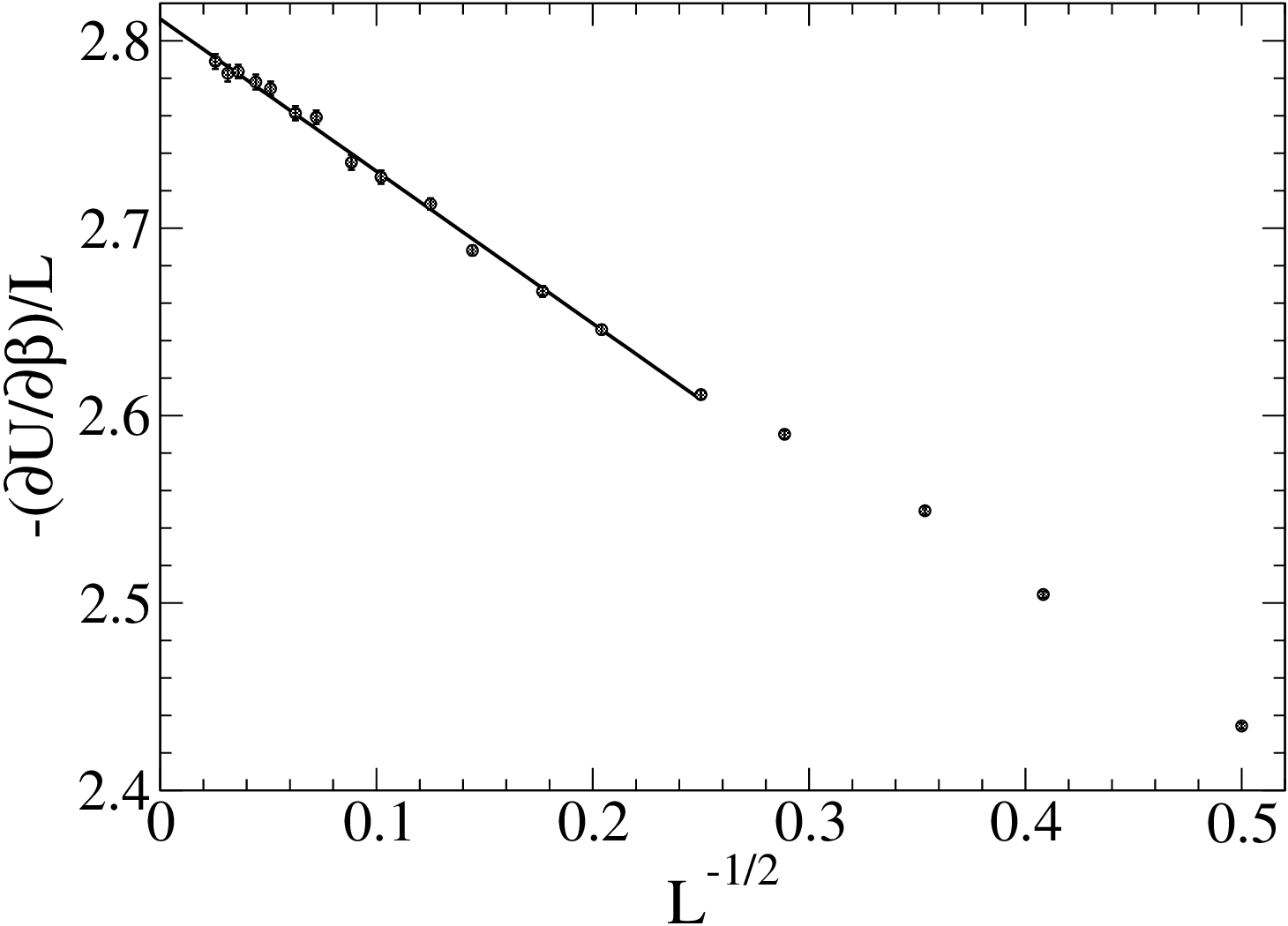}
\end{center}
\caption{The $(\partial U/\partial \beta)/L$ vs $L^{-1/2}$ plot for $\lambda=0.1$ and $U=2$. The straight line
represents the fit to $A+BL^{-1/2}$ within $16 \le L \le 1536$.}
\label{uatvas}
\end{figure}

In order to estimate correction--to--scaling exponents, first we are looking for quantities,
which can be well fit over a wide range of sizes to the ansatz of the form $A + B L^{-\omega}$, 
including only a single correction exponent $\omega$. Obviously, the most serious estimation is
possible at $\lambda=0.1$, where the data up to $L=1536$ are available.
We have find that $(\partial U/\partial \beta)/L$ data at $\lambda=0.1$ and $U=2$
can be fairly well fit to this ansatz within $L \in [L_{\mathrm{min}},1536]$ for $L_{\mathrm{min}} \ge 16$.
These fits give $\omega = 0.470(27)$ with $\chi^2/\mathrm{d.o.f.}=1.23$ at $L_{\mathrm{min}}=16$,
 $\omega = 0.497(38)$ with $\chi^2/\mathrm{d.o.f.}=1.25$ at $L_{\mathrm{min}}=24$ and
 $\omega = 0.546(52)$ with $\chi^2/\mathrm{d.o.f.}=1.19$ at $L_{\mathrm{min}}=32$.
These values of $\omega$ are close to $1/2$. Intuitively, the exact value is expected to
be a simple rational number, since all known critical exponents of the 2D Ising universality class
are such numbers. Thus, the leading correction--to--scaling 
exponent in the scalar 2D $\varphi^4$
model can be just $\omega=1/2$. The actual $-(\partial U/\partial \beta)/L$ plot depending on $L^{-1/2}$
is shown in Fig.~\ref{uatvas}. This plot is approximately
linear within the whole range of sizes $4 \le L \le 1536$. The fit with fixed exponent $\omega=1/2$
is fairly good within  $16 \le L \le 1536$. This fit with $\chi^2/\mathrm{d.o.f.}=1.22$ is shown in
Fig.~\ref{uatvas} by straight line.

A reasonable explanation of these results is such that~(\ref{eq:uatv}) contains a term with the exponent $1/2$,
which is the leading term  within $16 \le L \le 1536$, at least, for the actual parameters $\lambda = 0.1$ and $U=2$.
According to the analytical arguments in Sec.~\ref{sec:analytical}, a correction term
with exponent $3/4$ exists in the two--point correlation function. Thus, it is expected in~(\ref{eq:chi}) -- (\ref{eq:uatv}), as well. 
As explained in Sec.~\ref{sec:analytical}, extra correction terms with smaller exponents are also possible.
The current analysis provides an evidence for such a correction with exponent $1/2$.
According to the predictions of~\cite{K_Ann01}, a correction term with exponent $1/4$ is also expected.
Our analysis of these data does not provide 
any evidence for such a correction. However, there is no contradiction with this conception, if we assume that the 
amplitude of the latter correction term is relatively small. In this case the behavior in Fig.~\ref{uatvas} 
should be changed for large enough lattice sizes to the $\sim L^{-1/4}$ asymptotic convergence.

This scenario is supported by the  $(\partial U/\partial \beta)/L$ data at $\lambda=0.1$ and $U=1.1679229 \approx U^*$.
These data are not well described by $A + B L^{-\omega}$, but can be quite well fit to a refined
ansatz of the form $A + B L^{-\omega} + C L^{-2 \omega}$. The exponent $\omega=1/4$ would confirm the aforementioned scenario.
The data for  $\lambda=0.1$ and $U=1.1679229 \approx U^*$ are satisfactory well fit to this ansatz 
within $L \in [48,1536]$, yielding $\omega=0.275(79)$. The  $\chi^2/\mathrm{d.o.f.}$ of this fit
is $1.08$, which is the smallest value among all fits within $[L_{\mathrm{min}},1536]$, for which
the number of degrees of freedom exceeds the number of fit parameters (i.~e., for $L_{\mathrm{min}} \le 96$).
This result is well consistent with $\omega=1/4$. Other estimates are $\omega=0.399(22)$ for $L_{\mathrm{min}}=16$,
$\omega=0.436(46)$ for $L_{\mathrm{min}}=24$,
$\omega=0.386(59)$ for $L_{\mathrm{min}}=32$ and  $\omega=0.26(11)$ for $L_{\mathrm{min}}=64$ with
$\chi^2/\mathrm{d.o.f.}=1.24, 1.24, 1.23$ and $1.24$, respectively.
We have performed also fits within $L \in [L_{\mathrm{max}}/32,L_{\mathrm{max}}]$ with different
maximal lattice sizes $L_{\mathrm{max}}$. The results are $\omega=0.477(77)$ for $L_{\mathrm{max}}=768$,
$\omega=0.394(78)$ for $L_{\mathrm{max}}=1024$ and $\omega=0.275(79)$ for $L_{\mathrm{max}}=1536$
with $\chi^2/\mathrm{d.o.f.}=1.48, 1.40$ and $1.08$, respectively. 
These $\omega$ values tend to decrease when the lattice sizes are
increased, showing that $\omega$ can be as small as $1/4$.
Another possibility is that $\omega$ has a larger value,
closer to $\omega=0.399(22)$, provided by the wide--range
fit over $L \in [16,1536]$.

As an extra test, we have fit these $(\partial U/\partial \beta)/L$ data at $\lambda=0.1$ and $U=1.1679229 \approx U^*$
to the ansatz $A + B L^{-1/2} + C L^{- \omega}$, where one of the correction exponents is set equal to $1/2$, in agreement 
with the behavior in Fig.~\ref{uatvas}. The fit within $L \in [48,1536]$ is fairly good ($\chi^2/\mathrm{d.o.f.}=1.07$) 
and gives $\omega = 0.34(26)$. It is consistent with our previous estimations, although the error bars are larger.

The analysis of corrections to scaling contained in $\chi/L^{7/4}$ is a more difficult problem
than the actual analysis of the $(\partial U/\partial \beta)/L$ data, since the susceptibility
$\chi$ contains a constant background contribution. The necessity to consider
several correction terms makes the estimation of correction exponents ambiguous. 
Due to this reason, we have only performed some consistency tests with fixed exponents for $\chi$,
as described in Sec.~\ref{sec:fixtest}.

\subsection{Test of the Ising scenario}
\label{sec:fixtest}

We have fit our susceptibility data for $\lambda=0.1$ within $L \in [\bar{L} /8, 8 \bar{L}]$
at different values of $\bar{L}$, using the ansatz 
\begin{equation}
 \frac{\chi}{L^{7/4}} = a_0 + a_1 L^{-1} + a_2 L^{-7/4} + a_3 L^{-2}
\label{eq:Isingc}
 \end{equation}
in order to test the consistency of the coefficients $a_1$ and $a_2$ with the Ising scenario
discussed in Secs.~\ref{sec:ancorr} and~\ref{sec:preliminary}. Namely, if corrections to scaling have the same structure
as expected in the 2D Ising model, then~(\ref{eq:Isingc}) holds at $L \to \infty$  
with $U$--independent value of $a_2$. 
The results depending on $\bar{L}$ are collected
in Tab.~\ref{tabfit}. 
\begin{table}
\caption{The fit parameters $a_1$ and $a_2$ in~(\ref{eq:Isingc}) depending on the fit interval
$L \in [\bar{L} /8, 8 \bar{L}]$ for the $\chi/L^{7/4}$ data with $U=1.1679229 \approx U^*$ in Tab.~\ref{tab2}.
The values of $a_2$ for the data with $U=2$ in Tab.~\ref{tab1} are denoted by $a_2^*$,
and $\Delta a_2$ is the difference $a_2-a_2^*$. The values of $\chi^2/\mathrm{d.o.f.}$ of the fits 
are shown in columns No.~4 and 7 for $U=1.1679229$ and $U=2$, respectively.}
\label{tabfit}
\begin{center}
\begin{tabular}{|c|c|c|c||c|c|c|}
\hline
\rule[-2mm]{0mm}{7mm}
$\bar{L}$ & $a_1$ & $a_2$ &  $\chi^2/\mathrm{d.o.f.}$  & $a_2^*$ & $\Delta a_2$ & $\chi^2/\mathrm{d.o.f.}$ \\
\hline
64  & 0.239(26) & 59.16(69) & 3.99 & 30.31(46) & 28.84(82) & 3.06 \\
96  & 0.104(37) & 64.6(1.3) & 1.60 & 33.95(88) & 30.6(1.6) & 2.29 \\
128 & 0.065(45) & 66.5(2.0) & 1.43 & 37.9(1.4) & 28.6(2.4) & 0.96 \\
192 & 0.039(65) & 68.8(3.8) & 1.45 & 40.8(2.8) & 28.0(4.7) & 0.90 \\
\hline
\end{tabular}
\end{center}
\end{table}
As we can see, $a_1$ for $U \approx U^*$ tends to zero with increasing of the lattice sizes used
in the fit, i.~e., with increasing of $\bar{L}$. It can be, indeed, expected in the Ising scenario.
The coefficient $a_2$ is slightly varied with $\bar{L}$ for both $U=1.1679229 \approx U^*$ and $U=2$.
The difference $\Delta a_2$ between the values of $a_2$ in these two cases, however,
is rather stable and clearly inconsistent with zero. Thus, the Ising scenario, where $\Delta a_2 \to 0$
at $\bar{L} \to \infty$, is not confirmed.

We have performed also the tests at $\lambda=1$. In this case, the data are fairly well fit within
$L \in [8,384]$, yielding $a_1=0.036(10)$ and $a_2=8.10(26)$  with
$\chi^2/\mathrm{d.o.f.} = 1.26$ for $U=1.1679229 \approx U^*$, and $a_2= 5.11(18)$ with 
$\chi^2/\mathrm{d.o.f.} = 1.17$ for $U=2$. The fits within $L \in [12,384]$ yield
$a_1=0.028(18)$ and $a_2=8.37(59)$ with $\chi^2/\mathrm{d.o.f.} = 1.40$ for $U=1.1679229$,
and $a_2=5.75(40)$ with $\chi^2/\mathrm{d.o.f.} = 0.87$ for $U=2$. As we can see, the coefficient $a_1$ is marginally well
consistent with zero, whereas the values of the coefficient $a_2$ for $U \approx U^*$
and $U=2$ are inconsistent, as in the case of $\lambda=0.1$.

 If the singular ``long--distance'' terms in susceptibility~(\ref{eq:chis}) (see the discussion in Sec.~\ref{sec:ancorr}) 
 contain only integer correction--to--scaling exponents,
then $a_2$ comes from the analytical background contribution and, thus, must be $U$--independent.
Consequently, the failure in our consistency tests strongly suggests that theses singular terms contain
nontrivial corrections to scaling, described by non--integer correction--to--scaling exponents.
If the expansion of $\chi/L^{7/4}$ contains all positive integer powers of $L^{-1/4}$, then
both singular and analytical parts of susceptibility $\chi$ contribute to the coefficient at $L^{-7/4}$
and, therefore, this coefficient is $U$--dependent.
 
We have performed one more test of the Ising scenario, using the $\partial U/\partial \beta$ data 
at $U=1.1679229 \approx U^*$. According to this scenario, a non-vanishing term $\propto L^{-2}$ 
is always expected in~(\ref{eq:uatv}). Therefore we have fit these data to
\begin{equation}
 \frac{1}{L} \frac{\partial U}{\partial \beta} = A + B L^{-2} + C L^{-\omega}
\label{eq:isansatz}
 \end{equation}
to test how well the extra exponent $\omega$ is consistent with
an integer value, which is different from $2$, as it must be true if the Ising scenario holds.
The results for $\omega$ depending on the fit interval $L \in [L_{\mathrm{min}},1536]$ are 
collected in Tab.~\ref{tabistest}.
\begin{table}
\caption{The fitted values of the exponent $\omega$ in~(\ref{eq:isansatz}) depending on
$L_{\mathrm{min}}$ for fits within $L \in [L_{\mathrm{min}},1536]$.
The quality of the fits is characterized by quantities $\chi^2/\mathrm{d.o.f.}$ ($\chi^2$ per degree of freedom)
and $Q$ (goodness of the fit).}
\label{tabistest}
\begin{center}
\begin{tabular}{|c|c|c|c|}
\hline
\rule[-2mm]{0mm}{7mm}
$L_{\mathrm{min}}$ & $\omega$ & $\chi^2/\mathrm{d.o.f.}$ & $Q$  \\
\hline
6  & 1.188(15) & 4.11 & 0.00000074  \\
8  & 1.299(25) & 1.67 & 0.066   \\
12 & 1.373(48) & 1.50 & 0.123   \\
16 & 1.418(81) & 1.60 & 0.099  \\
24 & 1.40(14)  & 1.78 & 0.066  \\
\hline
\end{tabular}
\end{center}
\end{table}
The values of $\chi^2/\mathrm{d.o.f.}$, as well as the values of the goodness $Q$ of the fit~\cite{Recipes} in
Tab.~\ref{tabistest}, show that these fits have a rather low quality. In fact, only the fits with $Q>0.1$ are
normally accepted~\cite{Recipes}, so that only the fit with $L_{\mathrm{min}} =12$ is more or less acceptable, 
and $\omega=1.373(48)$ is the best estimate in Tab~\ref{tabistest}.
We have skipped the results for 
$L_{\mathrm{min}} > 24$, since these fits are not better
and have remarkably larger statistical errors. 
The low quality of the fits indicate that~(\ref{eq:isansatz}), probably, is not the correct
asymptotic ansatz. Moreover, the estimated values of $\omega$ and the best estimate $\omega=1.373(48)$ 
are inconsistent with any integer value. Thus, the Ising scenario is, again, not confirmed.

\section{Summary and conclusions}

Corrections to scaling in the scalar 2D $\varphi^4$ model have been studied
based on non-\linebreak perturbative analytical arguments and Monte Carlo analysis. 
The analytical results are based on certain scaling assumptions and the theorem
proven in Sec.~\ref{sec:analytical}. Important conditions of the theorem have been
numerically tested and confirmed in Sec.~\ref{sec:t}, using the Monte Carlo
results described in Secs.~\ref{sec:MCs} and~\ref{sec:crp}.

Our analysis supports the finite--size 
corrections near criticality, representable by an expansion of a correction 
factor in powers of $L^{-1/4}$. Following~\cite{CGNP11}, we allow that some of high--order expansion
terms in the scalar 2D lattice $\varphi^4$ model can be modified to include logarithmic factors. 
Analytical arguments show the existence of corrections
with the correction--to--scaling exponent $3/4$. A brief review of finite--size scaling relations 
and preliminary MC analysis of the data are provided in Secs.~\ref{sec:ancorr} -- \ref{sec:preliminary}.
The MC analysis of the $(\partial U/\partial \beta)/L$ data
in Sec.~\ref{sec:estex} provides an evidence that there exist corrections with the exponent $1/2$ and, 
very likely, also corrections with the exponent about $1/4$. 
The numerical tests in Sec.~\ref{sec:fixtest} clearly show that the structure of corrections to scaling 
in the 2D $\varphi^4$ model differs from that one expected in the 2D Ising model.

The overall behavior of the $(\partial U/\partial \beta)/L$ and $\chi / L^{7/4}$ data
can be interpreted in such a way that nontrivial corrections in the form of the expansion 
in powers of $L^{-1/4}$ generally exist, although corrections with $\omega_k<1$ 
in~(\ref{eq:chi}) and~(\ref{eq:uatv}) 
can be well detectable only for small values of $\lambda$, such as $\lambda = 0.1$, since the
amplitudes of these correction terms decrease with increasing of $\lambda$ and
approaching the Ising limit $\lambda \to \infty$. It naturally explains the fact that
some of the plots at $\lambda=10$, discussed in Sec.~\ref{sec:preliminary}, are almost linear, as it is
expected in the 2D Ising model.

Apart from corrections to scaling, we have estimated the critical coupling $\beta_c$ depending on $\lambda$
in Sec.~\ref{sec:crp} and have  discussed an interesting phenomenon that
the critical temperature ($1/\beta_c$) 
appears to be a non-monotonous function of $\lambda$. 

\section*{Acknowledgments}
 
The authors acknowledge the use of resources provided by the
Latvian Grid Infrastructure. For more information, please
reference the Latvian Grid website (http://grid.lumii.lv).\linebreak
R. M. acknowledges the support from the
NSERC and CRC program.


\begin{thebibliography}{100}


\bibitem{Amit}
D. J. Amit, \textit{Field Theory, the Renormalization Group, and Critical Phenomena},
World Scientific, Singapore, 1984.

\bibitem{Ma}
S. K. Ma, \textit{Modern Theory of Critical Phenomena},
W. A. Benjamin, Inc., New York, 1976.

\bibitem{Justin}
J. Zinn-Justin, \textit{Quantum Field Theory and Critical Phenomena},
Clarendon Press, Oxford, 1996.

\bibitem{Kleinert}
H. Kleinert, V. Schulte-Frohlinde, \textit{Critical Properties of $\phi^4$ Theories},
World Scientific, Singapore, 2001.

\bibitem{PV}
A. Pelissetto, E. Vicari, 
Phys. Rep. 368 (2002) 549--727.

\bibitem{K_Ann01}
J. Kaupu\v{z}s, 
Ann. Phys. (Berlin) 10 (2001) 299--331.

\bibitem{K2012x} J. Kaupu\v{z}s, Int. J. Mod. Phys. A 27, 1250114 (2012)

\bibitem{K2012} J. Kaupu\v{z}s, Canadian J. Phys. \textbf{9}, 373 (2012)

\bibitem{MHB86} A. Milchev, D. W. Heermann, K. Binder, J. Stat. Phys. \textbf{44}, 749 (1986)

\bibitem{TC90} R. Toral, A. Chakrabarti, Phys. Rev. B \textbf{42}, 2445 (1990)

\bibitem{MF92} B. Mehling, B. M. Forrest, Z. Phys. B \textbf{89}, 89 (1992)


\bibitem{KJJ06} R. Kenna, D. A. Johnston, W. Janke, Phys. Rev. Lett. \textbf{97}, 155702 (2006);
Erratum -- ibid \textbf{97}, 169901 (2006)

\bibitem{Kaupuzs06} J. Kaupu\v{z}s, Int. J. Mod. Phys. C \textbf{17}, 1095 (2006)

\bibitem{YP02} H. Au-Yang, J. H. H. Perk, Int. J. Mod. Phys. B \textbf{16}, 2089 (2002)

\bibitem{CGNP11} Y. Chan, A. J. Guttman, B. G. Nickel, J. H. H. Perk, J. Stat. Phys. \textbf{145},
549 (2011)

\bibitem{CHP02} M. Caselle, M. Hasenbusch, A. Pelissetto, E. Vicari, J. Phys. A \textbf{35}, 4861 (2002)

\bibitem{Sokal} J. Salas, A. D. Sokal, J. Stat. Phys. \textbf{98}, 551 (2000)

\bibitem{Perk} W.P. Orrick, B. Nickel, A.J. Guttmann and J.H.H. Perk,
J. Stat. Phys. \textbf{102}, 795 (2001)

\bibitem{AF80} A. Aharony, M. E. Fisher, Phys. Rev. Lett. \textbf{45}, 679 (1980)

\bibitem{AF83} A. Aharony, M. E. Fisher, Phys. Rev. B \textbf{27}, 4394 (1983)

\bibitem{BF84} M. Barma, M. Fisher, Phys. Rev. Lett. \textbf{53}, 1935 (1984)

\bibitem{Hasenbusch} M. Hasenbusch, J. Phys. A: Math. Gen. \textbf{32}, 4851 (1999)


\bibitem{MC} M. E. J. Newman, G. T. Barkema, Monte Carlo Methods
in Statistical Physics, Clarendon Press, Oxford, 1999

\bibitem{KMR_2010} J. Kaupu\v{z}s, J. Rim\v{s}\=ans, R. V. N. Melnik,
Phys. Rev. E \textbf{81}, 026701 (2010).

\bibitem{KMR_2011}
J. Kaupu\v{z}s, J. Rim\v{s}\=ans, R. V. N. Melnik,
Ukr. J. Phys. 56, 845 (2011)

\bibitem{Recipes} W. H. Press, B. P. Flannery, S. A. Teukolsky,
W. T. Vetterling, Numerical Recipes -- The Art of Scientific Computing,
Cambridge University Press, Cambridge, 1989

\bibitem{Baxter} R. J.~Baxter,
\textit{Exactly Solved Models in Statistical Mechanics},
Academic Press, London, 1989.

\bibitem{HasRev} M. Hasenbusch, Int.~J.~Mod.~Phys. C {\bf 12}, 911 (2001).


\end{thebibliography}
\end{document}